\documentclass[journal=ancac3,manuscript=article,layout=twocolumn]{achemso}
\usepackage{amsmath,amssymb,amsfonts}
\usepackage[osf]{newpxtext}
\usepackage{newpxmath}
\usepackage[utf8]{inputenx}
\usepackage{graphicx}
\usepackage[dvipsnames]{xcolor}
\usepackage[pdftex]{hyperref}
\usepackage{braket}
\usepackage{xspace}
\usepackage{soul}

\DeclareMathAlphabet{\mathcal}{OMS}{cmsy}{m}{n}

\hypersetup{
	pdfauthor={Miriam Oliva},
	bookmarksnumbered=true,
	pdftitle={Carrier Recombination in Highly Uniform and Phase-Pure GaAs/(Al,Ga)As Core/Shell Nanowire Arrays on Si(111): Mott Transition and Internal Quantum Efficiency},
	colorlinks,
	citecolor=blue,
	linkcolor=blue,
	urlcolor=blue}

\author{Miriam Oliva}
\email{miriam.oliva@pdi-berlin.de}
\author{Timur Flissikowski}
\author{Michał Góra}
\affiliation[1]{Paul-Drude-Institut für Festkörperelektronik, \\Leibniz-Institut im Forschungsverbund Berlin e.~V., \\Hausvogteiplatz 5–7, 10117 Berlin, Germany}
\alsoaffiliation[2]{Present address: Swiss Federal Laboratories \\for Materials Science and Technology, \\Advanced Fibers, St. Gallen, Switzerland}
\author{Jonas Lähnemann}
\author{Jesús Herranz}
\affiliation[1]{Paul-Drude-Institut für Festkörperelektronik, \\Leibniz-Institut im Forschungsverbund Berlin e.~V., \\Hausvogteiplatz 5–7, 10117 Berlin, Germany}
\alsoaffiliation[2]{Present address: Microsoft Quantum Materials Lab, \\Kanalvej 7, 2800 Kongens Lyngby, Denmark}
\author{Ryan B. Lewis}
\affiliation[1]{Paul-Drude-Institut für Festkörperelektronik, \\Leibniz-Institut im Forschungsverbund Berlin e.~V., \\Hausvogteiplatz 5–7, 10117 Berlin, Germany}
\alsoaffiliation[2]
{Present address: Department of Engineering Physics, \\McMaster University, L8S 4L7 Hamilton, Canada}
\author{Oliver Marquardt}
\affiliation[1]{Weierstraß-Institut für angewandte Analysis und Stochastik, \\Mohrenstraße 39, 10117 Berlin, Germany}
\author{Manfred Ramsteiner}
\affiliation[1]{Paul-Drude-Institut für Festkörperelektronik, \\Leibniz-Institut im Forschungsverbund Berlin e.~V., \\Hausvogteiplatz 5–7, 10117 Berlin, Germany}
\author{Lutz Geelhaar}
\author{Oliver Brandt}
\affiliation[1]{Paul-Drude-Institut für Festkörperelektronik, \\Leibniz-Institut im Forschungsverbund Berlin e.~V., \\Hausvogteiplatz 5–7, 10117 Berlin, Germany}
\email{oliver.brandt@pdi-berlin.de}

\title[Carrier Recombination in Highly Uniform and Phase-Pure GaAs/(Al,Ga)As Core/Shell Nanowire Arrays on Si$\boldsymbol{\mathsf{(111)}}$: Mott Transition and Internal Quantum Efficiency]
{Carrier Recombination in Highly Uniform and Phase-Pure GaAs/(Al,Ga)As Core/Shell Nanowire Arrays on Si$\boldsymbol{\mathsf{(111)}}$: Mott Transition and Internal Quantum Efficiency}


\begin{document}


\begin{tocentry}
\centerline{\includegraphics[width=\textwidth]{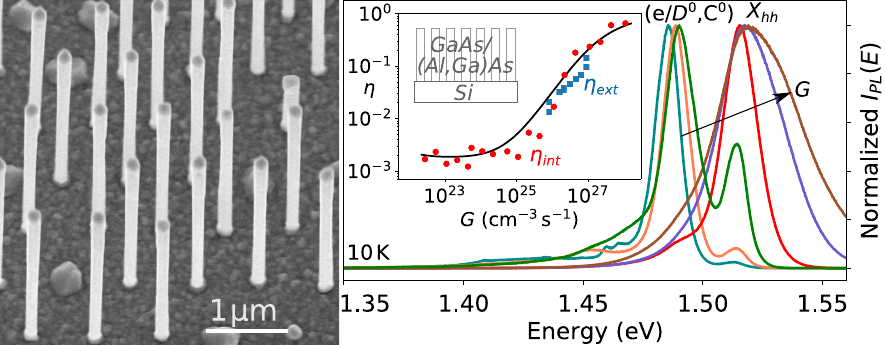}}
\end{tocentry}

\begin{abstract}

\textbf{\small
GaAs-based nanowires are among the most promising candidates for realizing a monolithical integration of III-V optoelectronics on the Si platform. To realize their full potential for applications as light absorbers and emitters, it is crucial to understand their interaction with light governing the absorption and extraction efficiency, as well as the carrier recombination dynamics determining the radiative efficiency. Here, we study the spontaneous emission of zincblende GaAs/(Al,Ga)As core/shell nanowire arrays by µ-photoluminescence spectroscopy. These ordered arrays are synthesized on patterned Si$\mathbf{(111)}$ substrates using molecular beam epitaxy, and exhibit an exceptionally low degree of polytypism for interwire separations exceeding a critical value. We record emission spectra over more than five orders of excitation density for both steady-state and pulsed excitation to identify the nature of the recombination channels. An abrupt Mott transition from excitonic to electron-hole-plasma recombination is observed, and the corresponding Mott density is derived. Combining these experiments with simulations and additional direct measurements of the external quantum efficiency using a perfect diffuse reflector as reference, we are able to extract the internal quantum efficiency as a function of carrier density and temperature as well as the extraction efficiency of the nanowire array. The results vividly document the high potential of GaAs/(Al,Ga)As core/shell nanowires for efficient light emitters integrated on the Si platform. Furthermore, the methodology established in this work can be applied to nanowires of any other materials system of interest for optoelectronic applications.
}  
\end{abstract}

\section{Introduction}
Planar group-III arsenide heterostructures enable optoelectronic devices such as vertical cavity surface emitting lasers in the near infrared spectral range. However, the monolithic integration of these GaAs-based devices on the Si platform remains an issue that has resisted a solution for decades despite considerable international research efforts.\cite{Fang_1990b,Bolkhovityanov_2008,Du_2022} The high lattice mismatch between GaAs and Si results in threading dislocation densities in excess of $10^7\,$cm$^{-2}$, and the high thermal mismatch leads to wafer bowing and cracks. Threading dislocations act as nonradiative defects, and their presence seriously degrades the internal quantum efficiency of any GaAs-based light emitter on Si. 

Both of these apparently insurmountable obstacles cease to be issues for GaAs-based nanowires (NWs) on Si$(111)$. Due to the limited footprint of these nanostructures, wafer bowing and cracks cannot occur. Furthermore, threading dislocation arms bend toward the NW surfaces close to the heterointerface,\cite{Hersee_2011} resulting in the complete absence of threading dislocations and the associated nonradiative recombination in GaAs NWs. Additionally, the high geometric aspect ratio of NWs offers a potentially more efficient absorption\cite{Heiss_2014} and extraction\cite{Hauswald_2017} of light when compared to planar structures. Consequently, the monolithic integration of III-V NW devices on Si substrates has attracted considerable attention, and has been successfully demonstrated by several groups \cite{Kikuchi_2004,Sekiguchi_2008a,Svensson_2008,Wei_2009,Tomioka_2010,Chuang_2011,Dimakis_2012,Krogstrup_2013,Borg_2014,Wu_2014,Koester_2015,Mayer_2016,Herranz_2020}.

GaAs NWs can be synthesized on Si(111) by both metal-organic vapor phase epitaxy (MOVPE) \cite{Martensson_2004,Roest_2006,Kang_2010,Huang_2010} and molecular beam epitaxy (MBE) \cite{Ihn_2007,Ihn_2007a,Mohseni_2007,Cirlin_2010,Uccelli_2011}. Most frequently, the vapor-liquid-solid (VLS) growth mode \cite{Wagner_1964} is employed, using either Au or Ga as the metal particles providing supersaturation, but growth in the vapor-soild (VS) growth mode has also been demonstrated. \cite{Tomioka_2010,Ruhstorfer_2020a,Ajay_2022}  In any case, the NWs grow preferentially along the $[111]$B direction, and are, regardless of growth technique and growth mode, prone to polytypism, i\,e., the stacking sequence along the length of the nanowire may change spontaneously between that of the zincblende or the wurtzite crystal structure.\cite{Spirkoska_2009,Plissard_2010,Heiss_2011a,Jahn_2012,Corfdir_2013,Corfdir_2016,Lin_2017,Senichev_2018} 

Crystallographically, polytypism manifests itself in stacking defects, namely, twin boundaries due to rotational twinning in the zincblende structure, and one of various types of stacking faults in the wurtzite structure. These defects may occur in an irregular sequence, and also longer segments of either crystal structure are possible. Electronically, these defects matter since the band alignment between wurtzite and zincblende GaAs is of type II.\cite{Spirkoska_2009,Kusch_2012,Kusch_2014,Senichev_2018} Thus, in the presence of stacking defects, spatially indirect, sub-bandgap transitions can occur between electrons and holes at the interfaces of the polytypes.\cite{Spirkoska_2009,Jahn_2012,Lin_2017,Senichev_2018} If the density of stacking defects is sufficiently high, the emission spectra of the NWs are dominated by these parasitic transitions, completely masking the intrinsic optical properties of the GaAs NWs. 

Several studies attempted to establish a correlation between microstructure and the transitions observed in single GaAs NWs.\cite{Perera_2008,Spirkoska_2009,Heiss_2011a,Jahn_2012,Ahtapodov_2012,DeLuca_2013,Loitsch_2015a,Lin_2017,Senichev_2018,Azimi_2021}. In some cases, single NWs were found exhibiting only a few stacking defects, thus displaying spectra dominated by the transitions typically observed in GaAs$(001)$ layers, particularly, the free exciton at higher excitation densities. Much more rare, but mandatory for any applications are ordered GaAs NW arrays with this desirable property, since any of these applications demand a deterministic and repeatable emission wavelength.

\citet{Joyce_2008} reported photoluminescence (PL) spectra dominated by the free exciton with a full-width at half-maximum of $7$\,meV for random GaAs/(Al,Ga)As core/shell NW ensembles synthesized by MOVPE on GaAs$(111)$B substrates. This group used the exceptionally high phase purity of their GaAs NW ensembles for detailed optical studies in subsequent work, deriving, for example, a value for the Mott density in GaAs NWs.\cite{Yong_2012} Also using GaAs$(111)$B substrates, \citet{Yang_2017} obtained even narrower exciton lines ($3$--$5$\,meV depending on Al content) for site-controlled GaAs/(Al,Ga)As nanomembranes (NMs) fabricated by MBE. The authors of the latter work demonstrated the high uniformity of their arrays by comparing PL spectra of single NMs with those of the entire array, which were found to be essentially identical.    

An even more important figure-of-merit for optoelectronic applications is the minority carrier lifetime $\tau$ or the internal quantum efficiency $\eta_\text{int}=\tau/\tau_r$, where $\tau_r$ is the radiative lifetime. Note that the stacking defects occurring in polytypic GaAs NWs induce spatially indirect radiative transitions of axially localized electron-hole pairs, and thus do not decrease, but rather tend to increase both $\tau$ and $\eta_\text{int}$. In contrast, the growth mode may play a crucial role, as shown by \citet{Breuer_2011} in their comparison of Au- and Ga-assisted GaAs/(Al,Ga)As NWs on Si$(111)$. Absolutely essential for obtaining high values for $\tau$ and $\eta_\text{int}$ is an efficient surface passivation because of the high surface recombination velocity of the bare GaAs$(110)$ sidewall surfaces.\cite{Demichel_2010,Chang_2012} The most convenient and effective passivation technique consists in the radial overgrowth of the GaAs core with (Al,Ga)As.\cite{Noborisaka_2005,Titova_2006,Perera_2008,Demichel_2010,Breuer_2011,Chang_2012} Still, even with an (Al,Ga)As shell, a finite interface recombination velocity remains, and the configuration of the passivating shell has been shown to affect $\tau$ in various ways.\cite{Jiang_2013,Kupers_2019} 

Except for the few facts summarized above, we actually know very little about the  mechanisms determining $\tau$ and $\eta_\text{int}$ in GaAs/(Al,Ga)As NWs. This statement applies to both radiative and nonradiative contributions, and in particular their dependence on excitation level. A thorough understanding of the carrier recombination dynamics in phase-pure GaAs/(Al,Ga)As NWs is essential for judging their eligibility for optoelectronic applications.     

In the present work, we use µ-PL spectroscopy to study the carrier recombination dynamics in ordered arrays of zincblende GaAs/(Al,Ga)As core/shell NWs grown on Si$(111)$ by MBE. The PL spectra of these NW arrays resemble those of the state-of-the-art NW and NM ensembles on GaAs$(111)$B and are essentially free of transitions induced by stacking defects down to the lowest excitation levels. Combining complementary spectroscopic experiments and simulations, we extract the internal quantum efficiency at $10$\,K for excitation levels spanning over more than five orders of magnitude, estimate the extraction efficiency, and measure the thermal quenching of the internal quantum efficiency for selected excitation levels. Furthermore, we observe an abrupt Mott transition from exciton to electron-hole-plasma recombination, allowing us to determine the Mott density.

\section{Results and discussion}
\label{sec:results} 

\subsection{NW morphology and PL spectra}
\begin{figure*}
		\vspace*{-1cm}
    \includegraphics[width=0.8\textwidth]{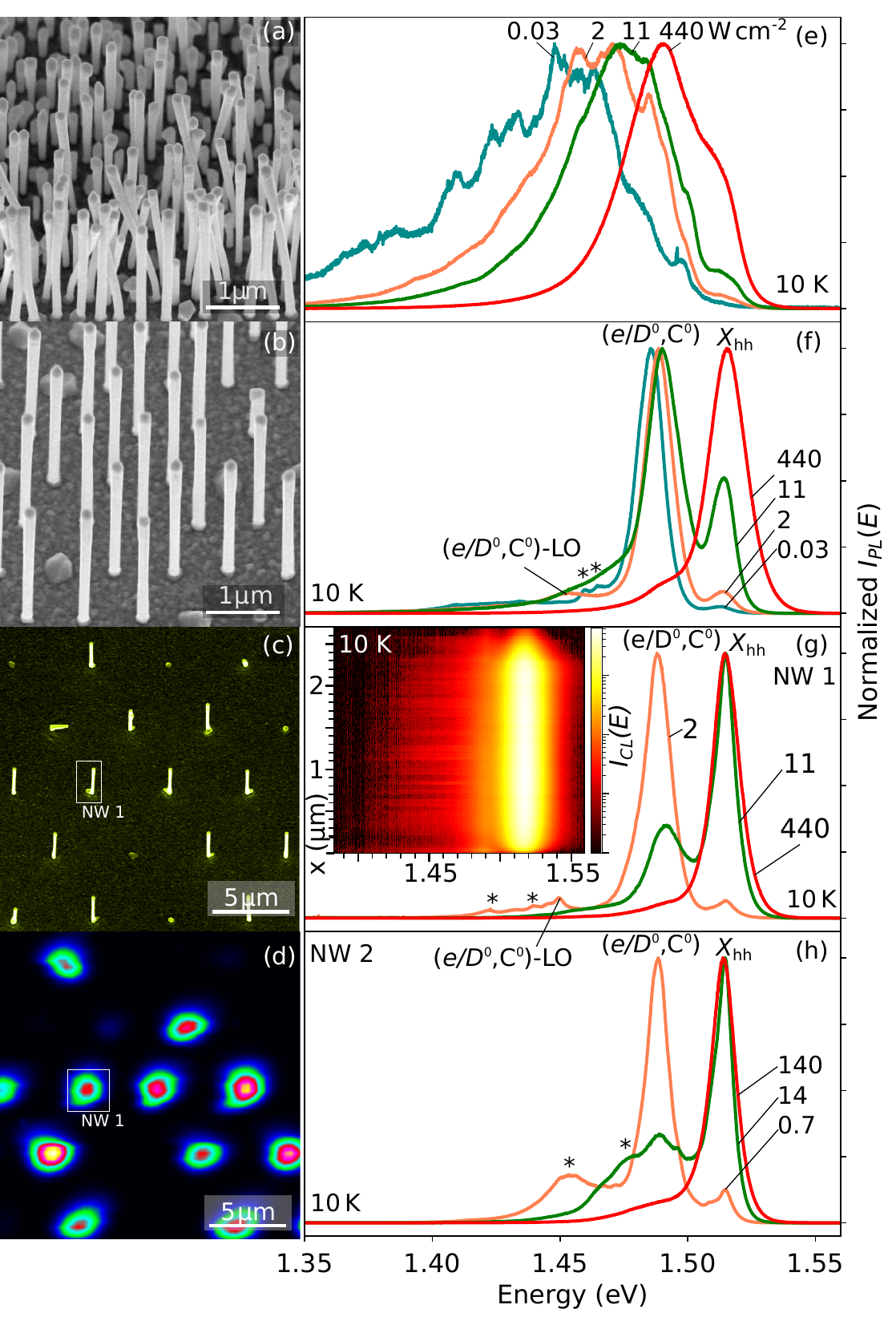}
    \caption{GaAs/Al$_{0.33}$Ga$_{0.67}$As core/shell NW arrays and their PL spectra. (a--c) Bird's eye ($15^{\circ}$) view secondary electron micrographs of regular NW arrays with $90\,$nm hole diameter and a pitch of $200$, $700$, and $5000$\,nm, respectively. (e--g) Low-temperature ($10\,$K) µ-PL spectra of the corresponding NW field recorded with different excitation densities $I_L$ in W\,cm$^{-2}$. The spectra in (g) stem from the single NW labeled NW~$1$ in panel (c). The inset in (g) depicts a hyperspectral cathodoluminescence linescan of another single NW. (d) µ-PL map of the $X_{\text{hh}}$ line recorded with $I_L=40$\,W\,cm$^{-2}$ of the microfield depicted in (c). (h) Low-temperature ($10\,$K) µ-PL spectra of a single NW (NW~$2$) from a row of single NWs with a distance of $10$\,µm and a hole size of $90$\,nm. The two dominant lines in (f), (g), and (h) originate from the C-assisted transitions $(e/D^0$,C$^0)$ and the free exciton ($X_\text{hh}$) in zincblende GaAs. Transitions presumably associated with polytypism in the spectra depicted in panels (f--h) are labeled with an asterisk. 
} 
    \label{fig1} 
\end{figure*}

Figure~\ref{fig1} depicts bird's eye ($15^{\circ}$) view secondary electron micrographs as well as low-temperature (nominally $10\,$K, actually $\approx 25\,$K -- see the Supporting Information for details) continuous wave (cw) µ-PL spectra of different microfields containing GaAs/Al$_{0.33}$Ga$_{0.67}$As core/shell NW arrays (sample A). The selected microfields have the same hole size of $90$\,nm, but different pitches of (a,e) $200$, (b,f) $700$ and (c,g) $5000$\,nm (see Sec.~\nameref{sec:Exp} for the definition of hole size and pitch). 

The secondary electron micrograph displayed in Fig.~\ref{fig1}a reveals a dense array of NWs with different lengths. The majority of NWs does not exceed a length of $2$\,\textmu m, but some approach the intended length of $4$\,\textmu m. This observation is likely an effect of shadowing of the Ga flux in the dense NW array, as investigated both experimentally and theoretically for Au-assisted InAs \cite{Madsen_2013} and Ga-assisted GaAs NWs.\cite{Schroth_2019} The bending of the long NWs arises from electrostatic effects during the exposure of the NWs to the electron beam in the scanning electron microscope. 

Figures~\ref{fig1}b,c show that the vertical yield improves drastically for fields with a larger NW pitch. In both cases, the micrographs show NWs of about $4$\,\textmu m length and $160$\,nm diameter (the GaAs core is $100\,$nm-thick) in an ordered hexagonal arrangement. The vertical yield exceeds $90$\% and parasitic growth is almost absent. The µ-PL map displayed in Fig.~\ref{fig1}d, which corresponds to the microfield of panel c, shows that only completely developed NWs exhibit a notable PL intensity, while the rare parasitic growth or NW stumps do not emit any detectable PL. Among the long vertical NWs, 11 out of the 12 of those visible in Fig.~\ref{fig1}c emit with essentially equal intensity.  

The PL spectra in Fig.~\ref{fig1}e are representative for NW arrays with a pitch of $200$--$400$\,nm, and are characterized by a broad band exhibiting a pronounced finestructure at low excitation density and a continuous blueshift with increasing excitation density. These features are typical of GaAs/(Al,Ga)As NWs with a high degree of polytypism\cite{Spirkoska_2009,Heiss_2011a,Jahn_2012,Corfdir_2013,Corfdir_2016,Lin_2017}. The random stacking sequences in the NWs result in a band of spatially indirect transitions with an almost continuous energy variation \cite{Jahn_2012,Marquardt_2017}. Note that the peak energy of this band does not reach the band edge of GaAs even for the highest excitation density.

In contrast, the spectra of NW arrays with a pitch  $\geq 700$\,nm (Fig.~\ref{fig1}f) are dominated by two well-defined lines. The spectral position (accounting for strain) and relative intensity as a function of excitation density $I_L$ of these lines, as well as their uniform appearance from NW to NW and along the length of single NWs (see below), identifies them as originating from the transitions prevalent in high-quality epitaxial GaAs layers (see the spectra in the Supporting Information for comparison). The dominance of these two lines even at the lowest excitation densities is a strong indication that the NWs possess the zincblende structure with a very low density of stacking defects such as twin boundaries or thicker wurtzite segments with stacking faults. Specifically, the line at $1.483$--$1.491$\,eV that prevails for low excitation densities stems from transitions related to the unintentional incorporation of the shallow acceptor C into the NWs, namely, the donor-acceptor-pair [($D^0$,C$^0$)] transition at the lowest and the conduction-band-to-acceptor [($e$,C$^0$)] transition at higher excitation density.\cite{Ashen_1975,Pavesi_1994} The individual contributions are not spectrally resolved since the width of the band encompasses both of the underlying transitions. However, the blueshift of the band with increasing $I_L$ is partly due to the actual characteristic blueshift of the ($D^0$,C$^0$) transition \cite{Pavesi_1994}, and partly an apparent blueshift due to the ($e$,C$^0$) transition overtaking the ($D^0$,C$^0$) transition at higher excitation density ($I_L > 1$\,W\,cm$^{-2}$). In the following, we will therefore refer to this band as ($e/D^0$,C$^0$). The weak signals at about $1.450$ and $1.414$\,eV are due to the first and second longitudinal optical (LO) phonon replicas of the ($e/D^0$,C$^0$) transitions. 

For $I_L > 20$\,W\,cm$^{-2}$, the ($e/D^0$,C$^0$) band saturates altogether and a line at $1.511$--$1.514$\,eV starts to dominate the spectra. The spectral position of this line identifies it to be of excitonic origin. The slight blueshift with increasing excitation density suggests a participation of acceptor-bound [(C$^0$,$X_{\text{hh}}$)] and free excitons ($X_{\text{hh}}$), although their individual contributions are not resolved in the spectra shown in Fig.~\ref{fig1}f. Because of the uniaxial tensile strain exerted by the (Al,Ga)As shell (see Supporting Information), these excitons are of heavy-hole character.\cite{Marquardt_2017}

The full width at half maximum (FWHM) of the $X_{\text{hh}}$ line in Fig.~\ref{fig1}f amounts to about $7$--$8$\,meV. While being remarkably narrow compared to the values commonly observed for GaAs NWs \cite{Rudolph_2014,Geijselaers_2018} with the exception of Ref.~\citenum{Joyce_2008}, it is still rather broad compared to the $X_{\text{hh}}$ line of planar GaAs for which we observe an FWHM of $(1.6 \pm 0.4)$\,meV (see Supporting Information). With a pitch of $700$\,nm and a laser spot size of $(2 \pm 0.4)$\,\textmu m, the spectra collected from this NW array originate from about $20$ individual NWs. To examine whether the broadening is simply an ensemble effect, we record spectra from single NWs in arrays with a pitch of $5$\,\textmu m or larger such as depicted in Fig.~\ref{fig1}c. Figures~\ref{fig1}g,h show spectra of two of such NWs. The spectra of NW1 in Fig.~\ref{fig1}g are representative for most of the NWs and are essentially indistinguishably from the ensemble spectra depicted in Fig.~\ref{fig1}f. This result suggests a high NW-to-NW homogeneity due to the virtual absence of polytypism, which would result in spectra that drastically differ from NW to NW.\cite{Spirkoska_2009,Heiss_2011a,Jahn_2012,Lin_2017} However, in some NWs we do detect transitions associated with stacking defects in spectra recorded at low excitation densities, such as in those of NW2 shown in Fig.~\ref{fig1}h (labeled with an asterisk). Still, even in this case, the spectra are dominated by the ($e/D^0$,C$^0$) and $X_{\text{hh}}$ lines that are not resolvable in spectra of NWs with a high degree of polytypism as those displayed in Fig.~\ref{fig1}e. 

The inset of Fig.~\ref{fig1}g shows a cathodoluminescence (CL) hyperspectral linescan of another single NW. Evidently, both the C-assisted and the excitonic emission intensities are uniform along the entire length of the NW, in striking contrast to linescans taken from NWs with a high degree of polytypism.\cite{Lin_2017} For the present NWs, we were unable to detect any transitions associated to stacking defects, presumably because the excitation density in CL is high enough to easily saturate transitions from isolated stacking defects (cf.\ spectrum in Fig.~\ref{fig1}h taken at $140$\,W\,cm$^{-2}$). For all single NWs, the FWHM of the $X_{\text{hh}}$ line amounts to about $5$--$6$\,meV and is thus only insignificantly narrower than the corresponding transition in the ensemble spectra. This finding shows that the broadening of the $X_{\text{hh}}$ line is not an ensemble effect. Possible mechanisms for the broadening of the transitions from single NWs will be discussed below. 

The results above are representative for two categories of microfields. From those with a pitch $<700$\,nm we invariably record PL spectra characteristic of pronounced polytypism. In contrast, for all fields with a pitch $>700$\,nm (up to $10$\,µm) we obtain spectra that resemble those of bulk GaAs except for the larger spectral width of the individual lines, which we will discuss later in Sec.~\nameref{Sec:linewidth} in detail. 
We observe this behavior for three samples containing either GaAs/Al$_{0.33}$Ga$_{0.67}$As (sample A) or GaAs/Al$_{0.10}$Ga$_{0.90}$As (samples B and C) core/shell NW arrays. The very different degrees of polytypism are observed in both samples for microfields in immediate vicinity ($<1\,$mm) and are thus definitely caused by the different pitch of the fields. In fact, it has been reported that the preferred polytype depends on the NW density due to the shadowing of the Ga flux\cite{Schroth_2018,Schroth_2019}, with zincblende being favored for larger interwire separation. Shadowing is indeed evident from the short NWs in the field with $200$\,nm pitch (cf.\ Fig.~\ref{fig1}a). We also note that shadowing, in conjunction with the continuous rotation of the sample during growth, leads to a periodical change of the effective V/III ratio at the tip of the NW, which may induce an instability at the triple-phase boundary (between the liquid Ga droplet, the GaAs NW, and the vacuum) with respect to the nucleation of zincblende or wurtzite monolayers.  

In any case, the PL spectra of NW arrays with a pitch equal or larger than $700$\,nm are close to those expected for phase-pure zincblende GaAs, and thus offer the opportunity to study the intrinsic optical properties of GaAs NWs in the sense that polytypism does not significantly affect these properties. In the remainder of this work, we consequently focus on the analysis of the PL spectra of NW fields with this desirable property, specifically, fields with a pitch of  $700$ and a hole diameter of $90$\,nm from samples A and B (for brevity, we refer to these fields in the following as $700/90$ NW arrays). In particular, we are interested in the internal quantum efficiency of the NWs in these fields.

\subsection{Excitation-density dependence}

\begin{figure} 
\vspace*{-8mm}
\includegraphics[width=\columnwidth]{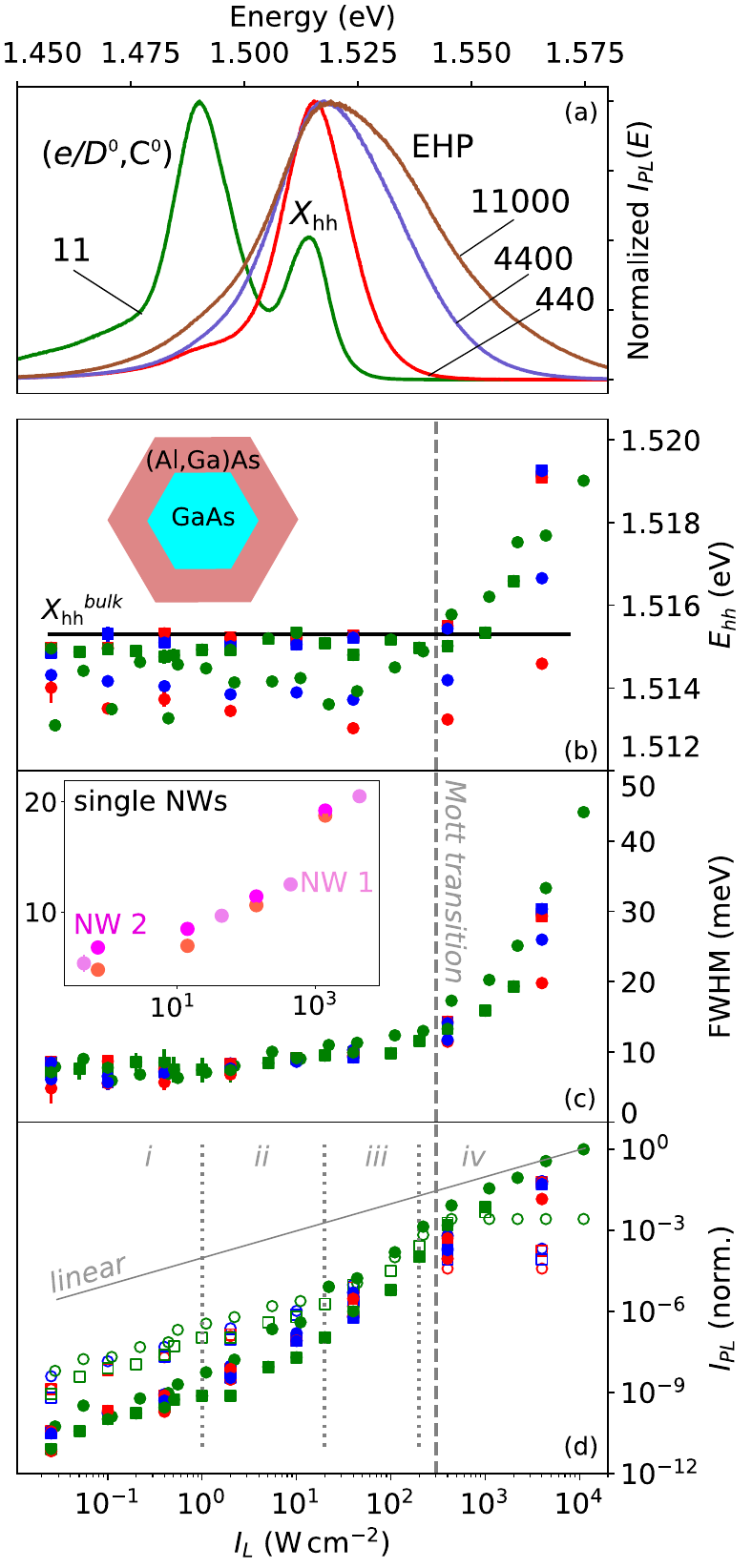}
    \caption{(a) Normalized $10$\,K µ-cw-PL spectra of the $700/90$ NW array from sample A, labeled with the respective $I_L$ in W\,cm$^{-2}$. (b) Spectral position $E_\text{hh}$, (c) FWHM, and (d) spectrally integrated PL intensity $I_{PL}$ of the $X_{\text{hh}}$ (full symbols) and the ($e/D^0$,C$^0$) (open symbols) transitions versus $I_L$. Circles (squares) correspond to measurements on sample A (B). Green, blue, and red data points refer to ensemble measurements on microfields with a pitch of $700\,$nm and hole diameters of, respectively, $90$, $70$, and $50\,$nm.  The inset in (b) depicts a schematic cross-section of the GaAs/(Al,Ga)As core/shell NWs under investigation. The inset in (c) shows the FWHM of the $X_{\text{hh}}$ line for three single NWs.  The solid and dashed lines in (b--d) are all guides to the eye.
    } 
    \label{fig2} 
\end{figure}

To get access to the internal quantum efficiency of the phase-pure NWs, we extend the range of excitation densities to which they are subjected. Figure~\ref{fig2}a shows $10$\,K cw-µ-PL spectra of the GaAs/Al$_{0.33}$Ga$_{0.67}$As core/shell NW array depicted in Fig.~\ref{fig1}b for low to high excitation densities. As discussed before, the spectra are dominated by the ($e/D^0$,C$^0$) band at low, and by the $X_{\text{hh}}$ line at intermediate excitation densitites. Upon even higher excitation densities, the $X_{\text{hh}}$ line abruptly blueshifts and broadens at its high-energy side. As discussed in detail below, this blueshift and broadening signifies a transition from excitonic to electron-hole-plasma (EHP) recombination accompanied by a progressive band filling.

These changes in the spectra are summarized quantitatively in Figs.~\ref{fig2}b--d depicting the spectral position $E_{\text{hh}}$, the FWHM, and the integrated intensity $I_{PL}$ of the $X_{\text{hh}}$ (full symbols) and the ($e/D^0$,C$^0$) (open symbols) transitions, respectively, versus the excitation density $I_L$. These quantities were extracted by line-shape fits of spectra measured on various microfields on samples A (circles) and B (squares). Regardless of sample, hole diameter, and pitch, the data are seen to follow the same trends.

\subsubsection{Emission energy: Band bending, strain, and Mott transition}
We next discuss the dependence of the $X_{\text{hh}}$ transition energy on excitation density and the composition of the passivating (Al,Ga)As shell. The spectral position of the $X_{\text{hh}}$ line shown in Fig.~\ref{fig2}b is seen to be constant for values of $I_L$ between $0.03$ and $300\,$W\,cm$^{-2}$. In this range of excitation densities, the $X_{\text{hh}}$ line is redshifted with respect to the $X_{\text{hh}}$ energy in unstrained GaAs (indicated by the horizontal line) by, on average, ($1.2 \pm 0.3$) and $(0.3 \pm 0.1)$\,meV for samples A and B, respectively. Such a redshift with increasing Al content in the shell has been observed by other groups as well. \citet{Hocevar_2013} attributed this shift to strain exerted on the GaAs core by the lattice-mismatched (Al,Ga)As shell, and substantiated this interpretation by calculations of the strain state. In contrast, \citet{Yang_2017} disregarded strain as the origin based on an absence of a shift in Raman spectra. Instead, these authors attributed the redshift to spatially indirect transitions between electrons and holes induced by a radial band bending in the NW core due to interface states formed by unintentionally incorporated O. \citet{Hocevar_2013} and \citet{Yang_2017} also observed an additional red-shift of the $X_{\text{hh}}$ transition energy for decreasing excitation densities. The former authors attributed this shift to the gradual screening of axial piezoelectric fields in conjunction with compositional fluctuations in the shell, while the latter ones invoked the filling of interfacial trap states and a corresponding modification of a band bending. Finally, \citet{Geijselaers_2018} observed an increasing redshift with increasing diameter, and assigned this finding to the presence of radial electric fields induced by residual acceptors with a concentration of $10^{15}$--$10^{16}$\,cm$^{-3}$. In the case of a redshift induced by an electric field, recombination has to occur between spatially separated electron-hole pairs with a transition energy shifted due to a radial Stark effect.\cite{Lahnemann_2016}

The radiative lifetime of exciton-polaritons in high-quality GaAs/(Al,Ga)As double heterostructures is on the order of a few ns at low temperatures.\cite{tHooft_1987,Rappel_1988a} The exciton decay times for Ga-assisted GaAs/(Al,Ga)As NWs, regardless of whether being fabricated by MBE or MOVPE, are close to this value.\cite{Perera_2008,Breuer_2011,Kang_2012,Hocevar_2013} An exciton in an electric field ionizes when the tunneling time of the electron in the Coulomb potential of the hole becomes significantly shorter than the recombination time. Full ionization occurs at a field corresponding to a potential drop of one Rydberg across the Bohr radius of the exciton.\cite{Blossey_1970} For GaAs, this field amounts to $3$\,kV/cm. For fields of this magnitude, recombination can only occur between electron-hole pairs spatially separated by the radial fields, thus being shifted to lower energies (analogously to the Franz-Keldysh effect in bulk crystals). In our case, the energy of the $X_{\text{hh}}$ line does not depend on $I_L$ for low and intermediate excitation densities, i.\,e., recombination remains excitonic down to the lowest excitation density of $30$\,mW\,cm$^{-2}$ (which is well within the range of excitation densities used, for example, in Refs.~\citenum{Hocevar_2013,Yang_2017,Geijselaers_2018}). Hence, the radial electric field in the GaAs core of our NWs is too small to dissociate excitons prior to their recombination. As shown in the Supporting information, this situation changes for NWs of larger diameter (and thus larger fields), in which recombination is no longer excitonic.

To obtain quantitative information of the band bending, the shell-induced strain and the associated radial piezoelectric fields,\cite{Boxberg_2012,Moratis_2016} we have performed two-dimensional simulations of our GaAs/(Al,Ga)As core/shell NWs using the Poisson solver of \textsc{nextnano}\texttrademark\ and the strain, piezopolarization, and eight-band $\mathbf{k \cdot p}$ modules of \textsc{SPHInX}.\cite{Boeck_2011,Marquardt_2014} The results are presented in the Supporting Information, with the main insights summarized as follows. First of all, radial electric fields are induced intrinsically by the lattice-mismatched shell, but are too small ($\ll 1$\,kV/cm) to dissociate excitons for the present Al content and thickness of the shell. Second, the radial electric fields caused by the residual background doping stay below the strength required for dissociating excitons up to a residual doping density of $1 \times 10^{16}$\,cm$^{-3}$. Third, the $\mathbf{k \cdot p}$ calculations show that the redshift with increasing Al content can be safely assigned to strain. In fact, and as also discussed in the Supporting Information, values close to the experimental ones are obtained already with a crude approximation, taking into account solely the uniaxial strain induced by the lattice mismatch between core and shell along the NW axis and the respective cross-sectional areas, while neglecting the different elastic properties of core and shell.  

The constant energy of the $X_{\text{hh}}$ line observed for both samples A and B for $I_L < 300\,$W\,cm$^{-2}$ is expected for free exciton transitions because of the charge neutrality of the exciton.\cite{Shah_1977a,Egorov_1977,Fehrenbach_1982} For higher excitation density, the continuum redshifts (band-gap renormalization) and eventually crosses the energy of the exciton and thus induces a Mott transition from excitonic to EHP recombination.\cite{Kilimann_1977,Fehrenbach_1982} In our case, we observe the exact opposite behavior: at $I_L \approx 300\,$W\,cm$^{-2}$, the transition energy abruptly blueshifts, and the $X_{\text{hh}}$ line broadens primarily at its high-energy side (cf.\ Figs.~\ref{fig2}a--c). 

The high-energy broadening is indicative of a progressive band filling due to the quasi-Fermi level for electrons entering the conduction band, creating a degenerate electron gas. The blueshift of the high-energy half-maximum thus appears to be analogous to the dynamic Burstein-Moss shift \cite{Dapkus_1970,Shah_1976,Moss_1980} observed in pump-and probe absorption spectroscopy, and has indeed the same origin, but is \emph{not} equivalent to it. In the absence of disorder as induced, for example, by heavy doping, the optical transitions in both absorption and emission obey strict \textbf{k} conservation. In absorption, this requirement of vertical transitions directly results in the blueshift of the absorption onset known as the dynamic Burstein-Moss shift. In emission, however, photoexcited holes rapidly relax toward the top of the valence band, and recombination can in principle only occur at $\mathbf{k} = 0$. Transitions at higher energies require the presence of unoccupied (acceptor-like) states whose spread in \textbf{k} space allows recombination with electrons at $\mathbf{k} > 0$,\cite{Zhang_1991,Zhang_1992} and these states can be provided intentionally or unintentionally in GaAs by incorporating shallow acceptors such as Be.\cite{Zhang_1991,Zhang_1992} 

For our nominally undoped NW arrays, the unintentionally incorporated C acts as shallow acceptor, and the pronounced high-energy broadening observed in Fig.~\ref{fig2}a suggests a comparatively high C concentration. In fact, for epitaxial GaAs layers with an extremely low background doping ($p<10^{13}$\,cm$^{-3}$), we observe the theoretically expected redshift of the transition (band-gap renormalization) with increasing excitation density (see the Supporting Information for details). Note that the band-gap renormalization is also visible for the NWs by the redshifting low-energy half maximum of the EHP band evident in Fig.~\ref{fig2}a.\cite{Zhang_1991,Shah_1977a} For the bulk-like layers (see Supporting Information), the Mott transition appears to be continuous, with excitons and the EHP coexisting over a certain range of excitation densities, but this effect may simply arise from the spatial nonuniformity of the carrier density induced by the finite excitation depth and carrier diffusion.\cite{Kappei_2005} In striking contrast, for the core/shell NWs under investigation, the sudden disappearance of the $X_{\text{hh}}$ line and the associated abrupt shift of the transition energy signify a discontinuous (at least in comparison to the bulk-like GaAs layers as discussed in the Supporting Information) Mott transition from excitonic to EHP recombination. In fact, the coupling of light to a NW array is modified such as that for sufficiently large interwire separations ($>600$\,nm), the NWs are excited uniformly along their entire length.\cite{Heiss_2014} The NW array under investigation with a pitch of $700$\,nm is thus a realization of an electronically bulk-like GaAs sample with nearly uniform excitation density, and is thus ideally suited for studying the Mott transition in bulk GaAs.

The excitation density of $300\,$W\,cm$^{-2}$ at which the Mott transition occurs in our GaAs/(Al,Ga)As core/shell NWs is comparable to that observed for the Mott transition in state-of-the-art GaAs/(Al,Ga)As double-heterostructures as shown in the Supporting Information. At the first glance, this seems to imply that the coupled exciton-free carrier lifetimes are comparable as well, but we need to take into account the fact that the absorptance of the NW array is generally higher than that of a planar layer.\cite{Heiss_2014} In later sections, we will estimate the actual carrier density corresponding to this excitation density, determine the internal quantum efficiency, and measure the actual exciton and EHP lifetimes.

\subsubsection{Emission linewidth: Random dopant fluctuations and band filling} 
\label{Sec:linewidth}
Next, we discuss the physical origins of the spectral width of the $X_{\text{hh}}$ line. For $I_L<2$\,W\,cm$^{-2}$, the FWHM of the $X_{\text{hh}}$ line is constant and amounts regardless of the Al content to values of $5$--$8\,$meV (see Fig.~\ref{fig2}c). In the inset of Fig.~\ref{fig2}c, we show that measurements on three different single NWs (NW 1 and 2 defined in Fig.~\ref{fig1} and an additional single NW from the same sample) yield values in the same range, indicating that the width of the transition is not governed by ensemble effects, as already indicated in the context of Fig.~\ref{fig1}. For comparable temperatures and excitation densities, the narrowest FWHM of GaAs/(Al,Ga)As $3$D nanostructure arrays were reported by \citet{Yang_2017} for GaAs nanomembranes. These authors measured an FWHM of $2$ ($4$)$\,$meV for membranes passivated with Al$_{0.10}$Ga$_{0.90}$As (Al$_{0.33}$Ga$_{0.67}$As) shells. In contrast to our results, the FWHM thus exhibited a notable dependence on the Al content in the shell. This fact was not commented upon by the authors, but a natural explanation follows from their previous interpretation of the transition energies. Due to the band bending, a comparatively large FWHM could result from a superposition of transitions with different energies, and as the band bending is supposed to increase with Al content, so would the FWHM of the transitions. 

However, this interpretation is not consistent with the experimental data obtained for the NWs investigated in the present work. As discussed above, the dependence of the transition energies on Al content and excitation density we have displayed in Fig.~\ref{fig2}b is only compatible with free exciton recombination shifted by strain exerted on the core by the shell, but not with spatially indirect transitions between radially separated electrons and holes as proposed in Refs.~\citenum{Geijselaers_2018} and \citenum{Yang_2017}. The same applies to the FWHM of this transition. If the FWHM were to arise from a superposition of spatially indirect transitions in a radial electric field, it would be expected to monotonically decrease with increasing excitation density due to screening. Furthermore, the FWHM should depend on NW diameter similarly to the transition energy as discussed in Ref.~\citenum{Geijselaers_2018}. However, Fig,~\ref{fig2}c shows that the FWHM \emph{increases} monotonically with excitation density, contrary to the expected behavior for spatially separated electron-hole pairs in radial electric fields. Moreover, we do not observe any change in the FWHM of the $X_{\text{hh}}$ line for NWs of different diameter, as documented in the Supporting Information.

An alternative explanation of the comparatively large FWHM of the transitions in our NWs is based on the fact that the residual C acceptors in the NW core are ionized in equilibrium even at cryogenic temperatures due to charge transfer to the surface. Solving the Poisson equation for our NW geometry (see the Supporting Information for details) reveals that this effect also applies to the core/shell NWs under investigation: the acceptors are fully ionized, but for the core diameters of the present NWs, the core remains fully depleted. The charge fluctuations created by the randomly positioned acceptor ions induce potential fluctuations that lead to a broadening of near band-edge transitions \cite{Schubert_1997}. In the present case, these fluctuations are particularly large due to the absence of free carriers, resembling the case of a perfectly compensated semiconductor as considered in Ref.~\citenum{Morgan_1965}. The FWHM of $5$--$8$\,meV observed for our samples then corresponds to acceptor concentrations on the order of $10^{15}$\,cm$^{-3}$, at the lower end of the values estimated in Ref.~\citenum{Geijselaers_2018}, and still consistent with our above estimate of the maximum concentration compatible with free exciton recombination. 

Above an excitation density of $300$\,W\,cm$^{-2}$, the FWHM is seen to steeply increase from $10$ to more than $40$\,meV. As discussed above, the low-energy broadening represents the redshifting continuum due to bandgap renormalization, while the high-energy broadening reflects the progressive filling of the conduction band and the recombination of electrons at $\mathbf{k} > 0$ with neutral acceptors. Note that the presence of neutral acceptors at high excitation densities is not in contradiction with the above assignment of the broadening to ionized acceptors at low excitation densities. It is well known that impurities ionized by compensating defects (which are here introduced by surface states) can be neutralized upon photoexcitation by capture of the respective photogenerated carrier species.\cite{Schmidt_1992} Obviously, the fraction of photoneutralized states depends on the ratio of the impurity and the photogenerated carrier density, with the majority of the impurities being in an ionized state at low excitation density, and in a neutral state at high excitation density. Correspondingly, the FWHM of the band-edge transition may be dominated by random dopant fluctuations in the fomer, and by band filling in the latter case.       

\subsubsection{Emission intensity: competition of radiative and nonradiative transitions and internal quantum efficiency} 

Figure~\ref{fig2}d shows the dependence of the integrated PL intensity $I_{PL}$ of the $X_{\text{hh}}$ (full symbols) as well as the ($e/D^0$,C$^0$) (open symbols) lines on excitation density $I_{L}$. Clearly, $I_{PL}$ does not increase linearly with $I_L$ over the whole range of excitation densities for either of these lines, but exhibits a more complex dependence on $I_{L}$. As we will discuss in the following, this dependence reflects a competition between three different recombination channels, the two radiative transitions giving rise to the $X_{\text{hh}}$ and the ($e/D^0$,C$^0$) lines, and nonradiative Shockley-Read-Hall (SRH) recombination. 

In excitation regime \textit{(i)}, the PL spectrum is dominated by the lowest-energy transition, the ($e/D^0$,C$^0$) line. The dependence of this line on $I_L$ is slightly sublinear, while the intensity of the $X_{\text{hh}}$ line increases slightly superlinearly. In excitation regime \textit{(ii)}, the $X_{\text{hh}}$ line progressively overtakes the ($e/D^0$,C$^0$) line, which starts to saturate in intensity. In excitation regime \textit{(iii)}, the $X_{\text{hh}}$ exhibits a strong superlinear increase in intensity. This increase cannot arise from the competition with the ($e/D^0$,C$^0$) line, since it persists also for the total intensity of both lines. Rather, this increase reflects the presence of a nonradiative channel, which is gradually overtaken by radiative recombination in this excitation regime, resulting in an increase of the internal quantum efficiency $\eta_\text{int}$.\cite{Ding_1992,Brandt_1995a,Brandt_1995b,Brandt_1996} Finally, in excitation regime \textit{(iv)}, the spectra are dominated by the EHP band, the intensity of which enters a linear dependence on $I_L$, reflecting that radiative recombination starts to dominate. Consequently, $\eta_\text{int}$ approaches unity in this regime. 

The behavior sketched above applies to both samples A and B despite the different Al contents in their shells. This finding suggests that the nonradiative process is not related to surface recombination, since the barrier height for sample A is drastically higher than for sample B. We would, in principle, also expect a dependence of the interface recombination velocity on the Al content,\cite{Kupers_2019} but it is certainly possible that such a dependence only sets in at higher Al content. Alternatively, 
point defects in the NW core may act as nonradiative SRH centers independent of the Al content of the shell. In any case, an increase of $\eta_\text{int}$ as observed in Fig.~\ref{fig2}d can originate either from a speedup of radiative recombination, a slowdown of the nonradiative one, or both effects acting together.

For quantitatively analyzing these data, we integrate over both emission lines to obtain the total radiative intensity $I_r$ as well as the overall internal quantum efficiency $\eta_\text{int}$ by dividing the total intensity by excitation density. Models suitable for fitting such data are invariably based on the quasi-equilibria established between the individual populations at a given temperature and excitation density (such as the Saha equation for excitons and free carriers) in the spirit of Refs.~\citenum{Pickin_1990,Ridley_1990,Brandt_1995a,Brandt_1995b,Brandt_1996,Brandt_1998a}, thus reducing a system of coupled rate equations to a single one that fully accounts for the variation of the spectrally integrated PL intensity with excitation density. 

For the present case of transitions involving excitons, free carriers, a shallow impurity, and a nonradiative center, we arrive, when summing over all radiative contributions, at an equation that is formally identical to either the simple model developed in Refs.~\citenum{Brandt_1995a,Brandt_1995b} (hereafter referred to as model I) or, when taking into account the SRH kinetics of the nonradiative process, at the model discussed in Ref.~\citenum{Brandt_1996} (hereafter referred to as model II). In both cases, the contribution of excitons is effectively taken into account by an additional term to the radiative recombination coefficient derived by the principle of detailed balance. Likewise, the contribution of the C-assisted transition replaces the usual monomolecular term proportional to the background carrier density, which is zero in the present case of NWs being depleted due to charge transfer of carriers to surface states.\cite{Geijselaers_2018} Explicit expressions for models I and II are given in the Supporting Information.

Within the frame of model I, the increase of $\eta_\text{int}$ as observed in Fig.~\ref{fig2}d originates from a speedup of radiative recombination due to a transition from mono- to bimolecular recombination. In the present case of predominantly excitonic recombination, such a transition naturally takes place for excitation densities close to the Mott transition, i.\,e., in regime \textit{(iii)}. Indeed, when taking into account the decrease of the exciton binding energy $E_x$ with increasing excitation density,\cite{Brandt_1998b,Snoke_2008,Sekiguchi_2017a} the Saha equation\cite{Saha_1921,Dresser_1968} \cite{}

\begin{equation}
\frac{n_x}{np} \propto \exp\frac{E_x(n)}{k_B T}
\end{equation}

\noindent
actually predicts that the fraction of free carriers over that of excitons increases with excitation density.\cite{Snoke_2008,Sekiguchi_2010} Here, $n$, $p$, $n_x$, $k_B$, and $T$ denote, respectively, the electron, hole, and exciton concentration, the Boltzmann constant, and the temperature. Hence, recombination close to the Mott transition may take an increasingly bimolecular character even at low temperatures. 

Within the frame of model II, an additional mechanism for an increase of $\eta_\text{int}$ emerges from the potential slowdown of nonradiative recombination that may occur for recombination via an SRH center that preferentially captures minority over majority carriers. In this case, the recombination rate may decrease by a factor up to $\tau_p/\tau_n$ from small to large signal excitation, where $\tau_n$ and $\tau_p$ are the capture times for minority and majority carriers, respectively.\cite{Brandt_1996} In the present case, however, this mechanism is effectively blocked by the depletion of the NWs and the corresponding vanishing background carrier density, resulting in a SRH recombination rate close to the large-signal case regardless of excitation density. 

The two considerations above are based on the assumption that the radiative and nonradiative recombination coefficients are constant and do not depend on the carrier density. This assumption is violated for degenerate semiconductors, for which the recombination coefficients decrease with increasing density and eventually enter a $1/n$ dependence, thus effectively reentering a monomolecular recombination regime at high excitation densities.\cite{Oliveira_1993,Kioupakis_2013,Sinito_2019a,Sinito_2019b} Likewise, nonradiative recombination slows down in the degenerate regime.\cite{Landsberg_1957,Heasell_1968,vonRoos_1977} However, the radiative and nonradiative rates in the degenerate regime can no longer be represented in terms of population densities multiplied by rate constants or coefficients,\cite{Oliveira_1993} with the consequence that the concept of detailed balance to account for the coupling of these populations cannot be applied. The resulting equation systems are only accessible numerically, and less suitable for fits to experimental data. For simplicity, we treat both the radiative and the nonradiative coefficients as constants, which means that we will overestimate them in the degenerate regime. Since both of them decrease with increasing density in the degenerate regime, their ratio (determining $\eta_\text{int}$ as well as $I_{PL}$) is expected to stay approximately constant, and we will see below that this is indeed the case when comparing these values with the actual lifetimes.  

\begin{figure}[!t] 
\includegraphics[width=\columnwidth]{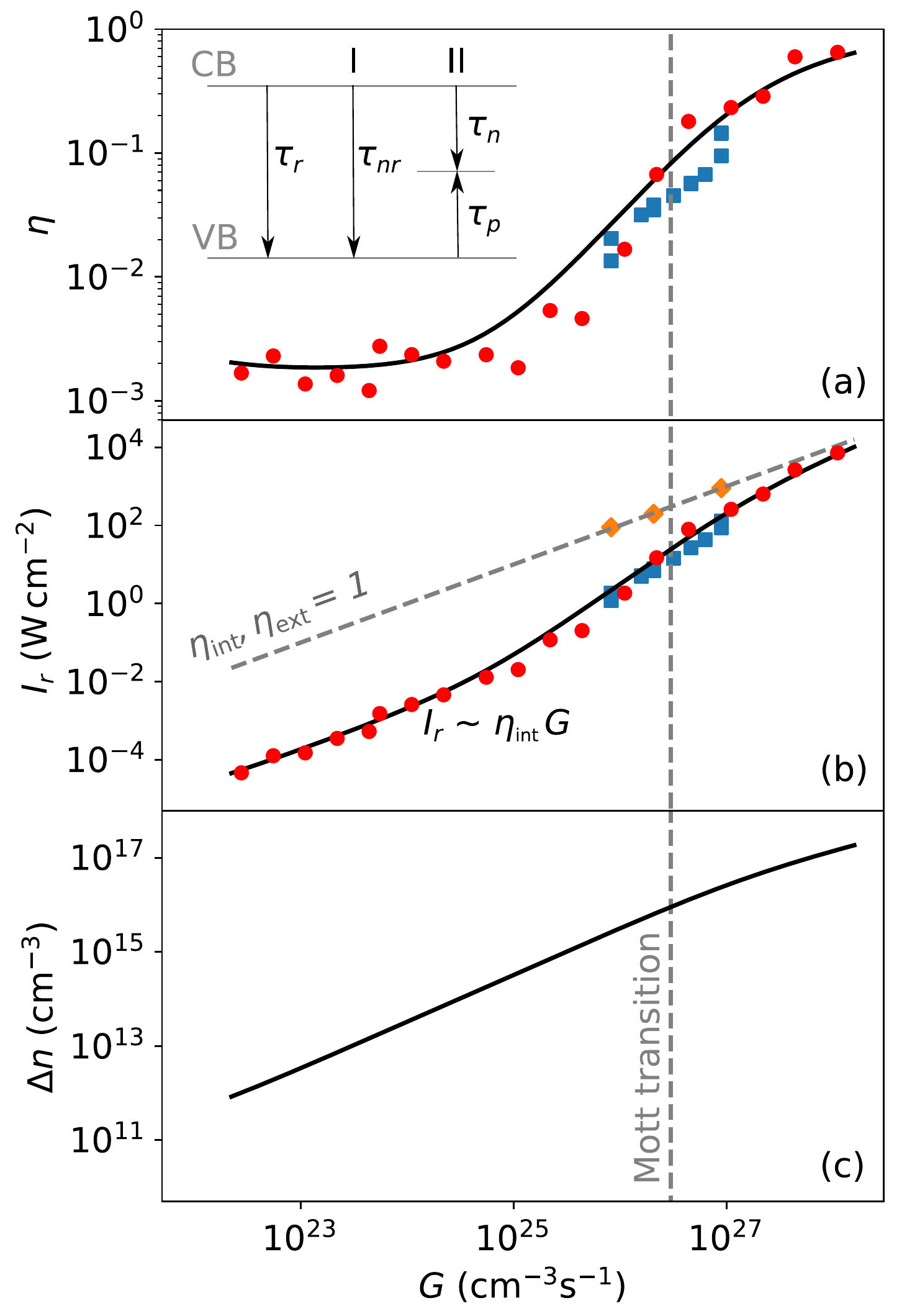}
    \caption{(a) Experimental values of the internal ($\eta_{\text{int}}=I_r/G$, circles) and external ($\eta_{\text{ext}}$, squares) quantum efficiency as a function of the generation rate $G$ scaled to match the relation between laser intensity and generation rate (see text). The solid line shows the fit of $\eta_{\text{int}}$ by model II. The inset displays a schematics of the radiative and nonradiative transitions considered in models I and II. (b) Total radiative PL intensity $I_r$ (circles and squares) obtained from the sum of the integrated intensities of the ($e/D^0$,C$^0$) and $X_{\text{hh}}$ lines shown in Fig.~\ref{fig2}a and the diffusively reflected laser intensity (diamonds) as a function of $G$. The dashed line indicates the case of $\eta_{\text{int}}$ and $\eta_{\text{ext}}$ equal to unity, and the solid line is the fit to the data represented by circles. (c) $\Delta n$ vs.\ $G$ as obtained by the fit of the data in (a,b). 
    } 
    \label{fig3}
\end{figure}

Figures~\ref{fig3}a,b show the dependence of the overall $\eta_\text{int}$ (as obtained by the spectrally integrated PL intensity divided by excitation density, and scaled to provide the best fit to the experimental data) and the total PL intensity on the generation rate $G$, respectively. The sketch in the inset of Fig.~\ref{fig3}a illustrates the transitions relevant in the context of models I and II. To actually compare the experiment to simulations, both the $y$ and $x$ axes are rescaled to the units used in the simulation. For the carrier generation rate $G =  I_{L}\,\alpha\,\hbar^{-1}\,\omega^{-1}$, we arrive at $G=10^{24}\,I_L$ by adopting an effective absorption coefficient of $2.5 \times 10^{5}\,$cm$^{-1}$ as computed by \citet{Heiss_2014} for GaAs NWs with a diameter equal to the one of the NWs under investigation in the present work. Note that this value is about $10\,\times$ higher than the one of bulk GaAs \cite{Sturge_1962}. The PL intensity is measured in arbitrary units, and our model thus includes a scaling factor to convert the PL efficiency (the PL intensity divided by the excitation density) to the internal quantum efficiency provided by the models. The best agreement is obtained when the point taken at highest excitation density is scaled to a value of $\eta_\text{int} = 0.65 \pm 0.1$.

The attraction of model I lies in its simplicity and the fact that it only has two free parameters, namely, the small-signal monomolecular radiative and nonradiative lifetimes $\tau_r^s$ and $\tau_{nr}$, respectively. In the present case, the former is given by the lifetime of the C-assisted transition $(b_A N_A)^{-1}$ with the recombination coefficient $b_A$ and the concentration of C acceptors $N_A$. The latter is the lifetime obtained from the full SRH expression for $p_0 \approx 0$ and $\Delta n \approx \Delta p$, i.\,e., $\tau_n + \tau_p$. Thanks to having only two free parameters, we obtain a unique fit as shown in Figs.~\ref{fig3}a,b returning values of $\tau_r^s = (19 \pm 2)$\,ns and $\tau_{nr} = (33\pm 5)$\,ps.

Model II has three free parameters, with which we found it difficult to achieve convergence and to obtain a unique parameter set. To facilitate convergence, we run the fit with the radiative lifetime fixed to certain preset values. The best fit is obtained with essentially identical parameters compared to model I: $\tau_r^s = 19$\,ns, $\tau_{n} = (12.5\pm 1)$\,ps, and $\tau_{p} = (20\pm 2)$\,ps. The fit also shows that $\Delta n \approx \Delta p $ except for the lowest generation rates. Hence, the total nonradiative lifetime varies only slightly from $37$ to $33$\,ps from low to high excitation density, and the fit is essentially indistinguishable from the one of model I, which is based on the assumption $\Delta n = \Delta p $ from the outset.

Figure~\ref{fig3}c shows the dependence of $\Delta n$ on the generation rate as derived from the best fit of model II to the experimental data. As expected, $\Delta n \propto G$ for small and $\Delta n \propto \sqrt{G}$ for large signal excitation. According to the fit, the excitation density of $300\,$W\,cm$^{-2}$, at which we observe the Mott transition to occur corresponds to a carrier density of $9 \times 10^{15}$\,cm$^{-3}$. This value is close to the theoretically expected one \cite{Edwards_1978} as well as the few available values for bulk GaAs.\cite{Shah_1977b,Sekiguchi_2017b} In the only previous study on GaAs NWs, the Mott transition was reported to be gradual and to take place between $2-4 \times 10^{16}$\,cm$^{-3}$.\cite{Yong_2012} Note that the nature of the Mott transition (i.\,e., continuous vs discontinuous) is not clear even for quantum wells with their inherently uniform excitation.\cite{Kappei_2005,Semkat_2009,Manzke_2012a,Guerci_2019b} It is worth noting that degeneracy of the conduction band of GaAs sets in at $(10 \pm 3)\times10^{15}$\,cm$^{-3}$ at a temperature of $(25 \pm 5)$\,K (see Supporting Information), i.\,e., band filling occurs indeed in parallel to the Mott transition.

\subsubsection{External quantum efficiency}
The maximum internal quantum efficiency derived in the previous section is $0.65$. Within the framework of the two models employed, we estimate the margin of error in this value to be not overly large, certainly less than $\pm 20$\%. However, the models themselves are based on many assumptions and simplifications, and it is thus prudent to ask whether the order of magnitude of $\eta_\text{int}$ is reasonable or not.

To answer this question, we adopt the method used by \citet{Amani_2015} to determine an absolute value of the external quantum efficiency $\eta_\text{ext}$ from the sample under investigation. Specifically, we perform a side-by-side comparison of the PL intensity of the $700/90$ NW array from sample C with the laser intensity reflected by a \textsc{Zenith Polymer}\textregistered\ diffuse reflector, possessing a nearly ideal Lambertian reflectance characteristics. The $\eta_\text{ext}$ of our NW array is then obtained by normalizing the spectrally integrated intensity of the PL from the NW array to that of the exciting laser, after properly correcting for the known reflectance of the \textsc{Zenith Polymer}\textregistered\ and the system response to account for the different wavelengths of the PL signal and the laser. Note that a confocal setup, as used for all previous experiments, does not easily allow such a comparison, for which we have conducted these experiments in a separate non-confocal setup as described in the Methods section. 

The diamonds in Fig.~\ref{fig3}b represent, by definition, an internal and external quantum efficiency of unity, as they are equal to the measured intensities on the \textsc{Zenith Polymer}\textregistered\ reflector, corrected for both the system response and the reflectivity of $84\%$ for the \textsc{Zenith Polymer}\textregistered\ thickness of $500$\,\textmu m used for this experiment. The squares display the ratio of the spectrally integrated PL intensity of our sample and the laser intensity measured side-by-side, i.\,e., the external quantum efficiency of this sample. Clearly, these data follow the same trend as those acquired with the confocal setup, but their absolute value is lower because the extraction efficiency is not unity even for deep-subwavelength NWs.\cite{Hauswald_2017} In the present case, the extraction efficiency as given by the ratio of these two different measurements amounts to $0.5 \pm 0.15$ on average. This value is in good agreement with theoretical work considering the effective diameter/wavelength ratio of $\omega d/c \approx 1.6$ for the present NWs.\cite{Maslov_2006} 

\begin{figure*}[!t]
\includegraphics[width=\textwidth]{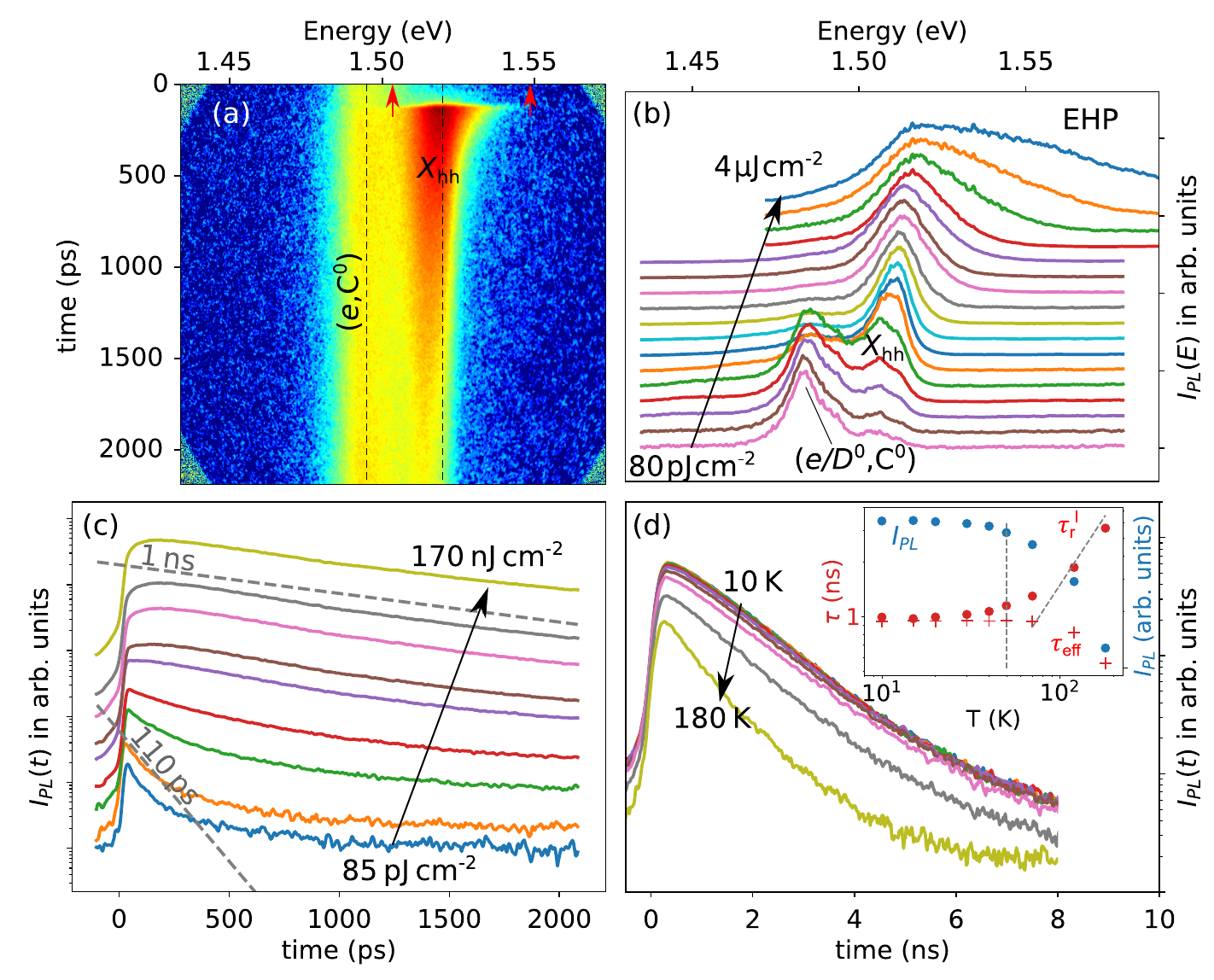}
    \caption{Carrier recombination dynamics upon pulsed excitation of the $700/90$ NW array from sample A. (a) Streak camera image taken at $10\,$K and a fluence of $400\,$pJ\,cm$^{-2}$, 
	showing the spectrally and temporally resolved $X_{\text{hh}}$ and ($e$,C$^0$) transitions. (b) PL spectra obtained from streak camera images in focus mode measured at $10\,$K, visualizing the evolution of the spectra from low to high fluences. (c) Spectrally integrated PL intensity transients of the $X_{\text{hh}}$ line [the red arrows shown in (a) indicate the integration range] obtained for different fluences as indicated in the figure. 
	(d) Spectrally integrated PL intensity transients measured at a fluence of $1.6\,$µJ\,cm$^{-2}$ and temperatures between $10$ and $180$\,K. The inset depicts the temperature dependence of $\tau_{\text{eff}}$, $\tau_r^\ell$, and $I_{PL}$ in a double-logarithmic representation. The vertical dotted line indicates $T=50\,$K, and the other is a guide to the eye displaying a $\tau_r \propto T^{3/2}$ dependence. } 
    \label{fig4} 
\end{figure*}

\subsection{Time-resolved PL experiments}
The comparison between the internal and external quantum efficiencies, derived by completely different means, shows that the ratio of the lifetimes returned by the fit to our experimental data (see Fig.~\ref{fig3}a,b) is quite accurate. However, in deriving the model, we have made several assumptions and performed many simplifications. Hence, the absolute values of these lifetimes are most likely not correct. To obtain these absolute values, and to confirm the interpretation of the cw-PL experiments, we investigate the $700/90$ NW array from sample A by time-resolved PL (TRPL) spectroscopy.

Figure~\ref{fig4}a shows a representative streak camera image taken at $10\,$K and a fluence of $400\,$pJ\,cm$^{-2}$. At this intermediate fluence, the $X_{\text{hh}}$ and the ($e$,C$^0$) lines are having comparable intensities. While the $X_{\text{hh}}$ line is seen to decay relatively fast, the intensity of the ($e$,C$^0$) line stays almost constant within the $2$\,ns time windows, reflecting the long lifetime typical for impurity-assisted transitions. Assuming an exponential decay as expected for the ($e$,C$^0$) transition, we obtain a decay time of $60$\,ns, corresponding (for a lattice temperature of $25$\,K) to an acceptor concentration of at most $1 \times 10^{16}$\,cm$^{-3}$ if this time corresponds to the actual radiative lifetime of this transition.\cite{Bebb_1972aa}

Figure~\ref{fig4}b depicts PL spectra extracted from streak camera images recorded in focus mode at $10$\,K and with different fluences as indicated in the figure. The evolution of the spectra is identical to the one observed in the cw-PL spectra shown in Figs.~\ref{fig1}f and \ref{fig2}a: at the lowest fluences, the spectra are dominated by the ($e/D^0$,C$^0$) line, which saturates at intermediate fluences, with the $X_{\text{hh}}$ line overtaking. At high fluences, the Mott transition takes place and band filling sets in, resulting in the blueshifted and asymmetrically broadened EHP emission.

Figure~\ref{fig4}c depicts spectrally integrated PL intensity transients of the $X_{\text{hh}}$ line at $10\,$K and different fluences as indicated in the figure. With increasing fluence, the effective lifetime $\tau_{\text{eff}}$ of the $X_{\text{hh}}$ line is observed to increase by one order of magnitude from about $0.11$ to $1$\,ns, reflecting the increase in $\eta_{\text{int}}(G) = \tau_{\text{eff}}(G)/\tau_r(G)$ with increasing generation rate extracted from the cw-PL measurements. The value of $1\,$ns is reached for fluences $\geq 9$\,nJ\,cm$^{-2}$, i.\,e., still well in the excitonic regime. This value is close to those reported for GaAs/(Al,Ga)As NWs \cite{Perera_2008,Breuer_2011,Kang_2012,Hocevar_2013} previously. The smooth transition from the excitonic to the degenerate EHP regime evidences the strong coupling of excitons and free carriers already at $10$\,K, resulting in a common lifetime for these states.\cite{Brandt_1998a} 

Figure \ref{fig4}d displays the temperature dependence of spectrally integrated PL intensity transients measured at a fluence of $1.6\,$µJ\,cm$^{-2}$. At this fluence, the instantaneous carrier density after the pulse amounts to $1.5\times10^{17}\,$cm$^{-3}$, corresponding to the highest carrier density created in the cw-PL experiments (cf.\ Fig.~\ref{fig3}c) and thus well above both the Mott transition and the crossover from a nondegenerate to a degenerate electron gas. The emission monitored thus stems from a degenerate EHP. Note that the transients are not scaled to allow a direct comparison of their intensities. For temperatures between $10$ and $50$\,K, the transients hardly change in both peak intensity and decay time, and thus integrated intensity. For higher temperatures, all of these quantities start to decrease.

The inset of Fig.~\ref{fig4}d displays the results of a quantitative analysis of the PL transients as function of temperature. Depicted are the effective lifetime $\tau_{\text{eff}}$ obtained from exponential fits to the transients, the large-signal radiative lifetime $\tau_r^\ell$ given (in arbitrary units) by the inverse of the transients maximum intensity,\cite{Sermage_1989,Deveaud_1991,Brandt_1996} and the integrated intensity $I_{PL}$. Clearly, for temperatures between $10$ and $50$\,K, all of these quantities are essentially constant, implying that $\tau_r^\ell \approx \tau_\text{eff}$ corresponding to an internal quantum efficiency $\eta_\text{int}$ close to unity. Note that $\tau_r^\ell$ is expected to be independent of temperature in a degenerate semiconductor, and to enter a $T^{3/2}$ dependence in the nongenerate regime\cite{Kioupakis_2013,Sinito_2019a,Sinito_2019b} For the present conditions, the crossover between these regimes occurs at around $80$\,K.

Finally, we can compare the small-signal lifetimes obtained from the fit of our simple recombination models to the cw-PL data (Figs.~\ref{fig2}a,b) and the decay times obtained by TRPL. As already stated above, the fit to the data returned values of $\tau_r^s = (19 \pm 2)$\,ns and $\tau_{nr} = (33\pm 5)$\,ps. The TRPL transients of the ($e/D^0$,C$^0$) and $X_{\text{hh}}$ lines taken at low fluences yield effective lifetimes of $60$\,ns and $110$\,ps. The former of these values is likely to correspond to the actual radiative lifetime of this transition, while the latter should be close to the actual nonradiative lifetime. The absolute value of these lifetimes is about a factor of three larger than those extracted from the fit, but their ratio---which determines the internal quantum efficiency---is essentially the same. In view of the assumptions and simplifications underlying the model, this result is very satisfactory. 


\subsection{Thermal quenching of the PL intensity}

\begin{figure}[!t]
  \includegraphics[width=\columnwidth]{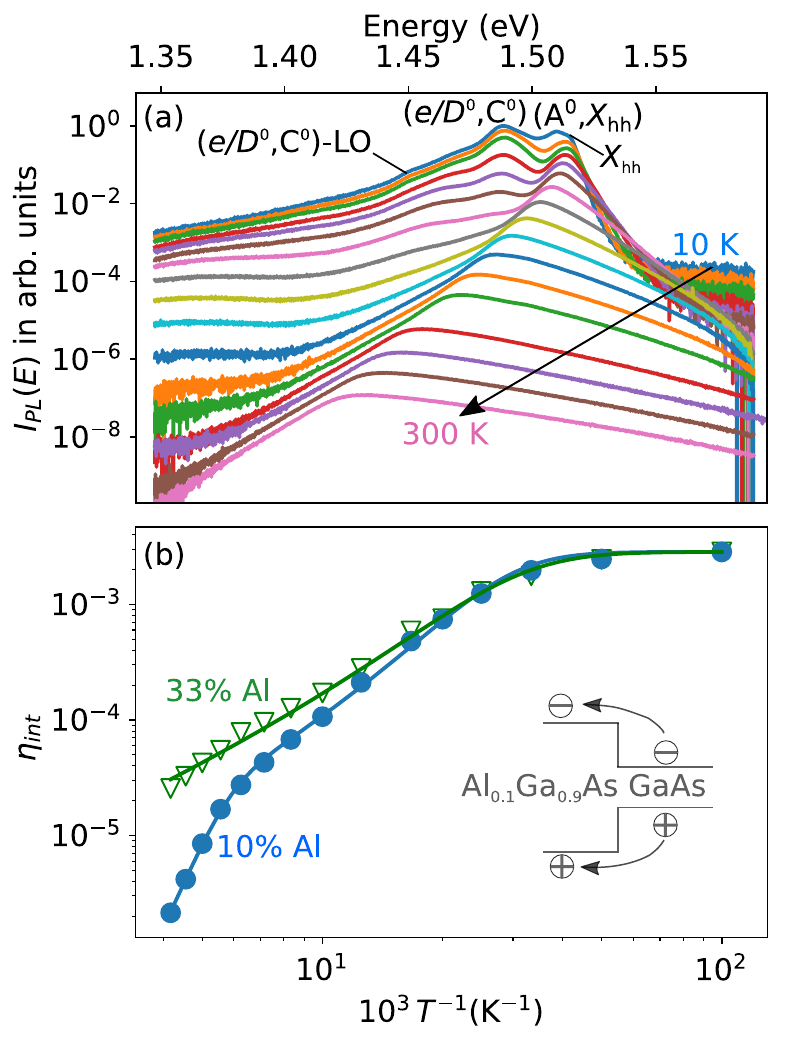}
    \caption{(a) Temperature-dependent cw-PL spectra of the $700/90$ NW array from sample A taken with $I_L = 10\,$W$\,$cm$^{-2}$ between $10$ and $300$\,K. (b) Double-logarithmic Arrhenius representation of the temperature-dependent internal quantum efficiency $\eta_\text{int}(T)$ of $700/90$ NW arrays from sample A (triangles) and B (circles). The solid lines are fits with Eq.~\ref{eq:arrhenius}. The inset schematically shows the bipolar thermionic emission of carriers across the energy barrier at the GaAs/Al$_{0.1}$Ga$_{0.9}$As interface.} 
    \label{fig5} 
\end{figure}

In the following, we analyze the temperature dependence of the cw-µ-PL data with a main focus on the effects of the Al content in the shell. Figure~\ref{fig5}a depicts cw-µ-PL spectra of the $700/90$\,nm NW array from sample A recorded at temperatures between $10$ and $300$\,K at an excitation density $I_L = 10$\,W\,cm$^{-2}$, i.\,e, in the monomolecular regime (excitation regime \textit{(ii)} in Fig.~\ref{fig2}d). For this set of measurements, we resolve the acceptor-bound exciton ($A^0$,$X_{\text{hh}}$) in addition to the ($e/D^0$,C$^0$) and $X_{\text{hh}}$ transitions at $10$\,K.  

The evolution of the spectra with temperature is identical to that observed for bulk GaAs.\cite{Grilli_1992} The individual transitions quench according to their binding energy,\cite{Bogardus_1968} and at temperatures $>200$\,K, only the band-to-band transition participates in radiative recombination. 

Figure~\ref{fig5}b displays a double-logarithmic Arrhenius representation of $\eta_\text{int}(T)$ measured for the $700/90$\,nm microfields of samples A and B. The data from both samples are fit by the familiar expression for a thermally activated quenching of the PL intensity or internal quantum efficiency \cite{Bimberg_1971,Kupers_2019} 
\begin{equation}
\eta_\text{int}(T) = \left[1+\sum_i a_i T^{3/2} \exp \left(-E_i/k_B T\right)\right]^{-1}
\label{eq:arrhenius}
\end{equation}
with the activation energies $E_i$ of the participating nonradiative channels, the prefactors $a_i$, the temperature $T$, and the Boltzmann constant $k_B$. The increase of the radiative lifetime with temperature in the bulk-like GaAs core is explicitly taken into account by the factor $T^{3/2}$.

The data obtained from sample A are fit well with a single activation energy amounting to $(6.6 \pm 0.5)$\,meV. This value is close to, but considering the error bar clearly larger than the exciton binding energy $E_X = 4.2$\,meV. In fact, the value represents an effective activation energy that subsumes over both the progressive dissociation of free excitons into free carriers and the thermal ionization of neutral C acceptors (with an energy of $E_X = 26$\,meV), resulting in the subsequent quenching of the ($e/D^0$,C$^0$) line. For a temperature up to $160$\,K, sample B exhibits the same behavior as sample A governed by an essentially identical activation energy. For temperatures higher than $160$\,K, however, $\eta_\text{int}$ of sample B declines sharply with a second activation energy of $(140 \pm 10)\,$meV. This value is very close to the band gap energy difference of Al$_{0.10}$Ga$_{0.90}$As and GaAs, indicating that this second nonradiative channel is related to the bipolar thermionic emission of carriers above the bandgap of the Al$_{0.10}$Ga$_{0.90}$As shell. \citet{Kupers_2019} recently arrived at the same conclusion for the thermal quenching of the PL intensity from GaAs/(In,Ga)As core/shell NWs. This additional process leads to a severe drop of $\eta_\text{int}$ at elevated temperatures. Extra\-polating the fits to $300$\,K, we obtain values of $2 \times 10^{-5}$ and $6 \times 10^{-7}$ for samples A and B, respectively. This result underlines the critical role of the shell for carrier recombination as already reported in previous studies.\cite{Jiang_2013,Kupers_2019}


\section{Summary and Conclusions}
\label{sec:summary}
In summary, we have presented a detailed spectroscopic investigation of ordered GaAs/(Al,Ga)As NW arrays on Si$(111)$. For a NW pitch equal or larger than $700$\,nm, we have found these arrays to exhibit a high phase purity and NW-to-NW uniformity, allowing us to quantitatively analyze systematic power-dependent, temperature- and time-dependent photoluminescence measurements. We have extracted several quantities of interest, such as the background acceptor concentration ($<10^{16}\,$cm$^{-3}$), the Mott density ($9\times 10^{15}\,$cm$^{-3}$), the extraction efficiency ($0.5$), and the internal quantum efficiency as a function of both carrier density and temperature (see below). To conclude, we put the values determined for the internal quantum efficiency at room temperature in broader context, and discuss the implications of our findings for applications of GaAs/(Al,Ga)As core/shell NWs in optoelectronics. 

At the first glance, the internal quantum efficiency of $2 \times 10^{-5}$ obtained for small-signal excitation at $300$\,K seems disappointingly low. However, our measurements truly represent a worst case scenario for (intentionally undoped) GaAs/(Al,Ga)As NWs. First of all, the NWs are fully depleted due to charge transfer to surface states (see the Supporting Information), i.\,e., the radiative rate at $300$\,K, where neither impurity-assisted nor excitonic transitions are active anymore, scales directly with the photogenerated carrier density, and has no monomolecular contribution. For the present excitation density of $I_L=10$\,W\,cm$^{-2}$, we have estimated an excess carrier density of $\Delta n \approx 3 \times 10^{14}$\,cm$^{-3}$ at $10$\,K (see Fig.~\ref{fig3}c), but this value will (due to thermally activated nonradiative channels) be certainly lower at $300$\,K. Consequently, the radiative lifetime $1/(B \Delta n)$ (with $B \approx 2 \times 10^{-10}$\,cm$^3$/s at $300$\,K) will be $\gg 10$\,\textmu s. Second, even in the total absence of nonradiative recombination in the bulk, the interface recombination velocity at GaAs/(Al,Ga)As interfaces is finite. For a NW diameter of $100$\,nm, the data compilation in Fig.\,2 of Ref.~\citenum{Breuer_2011} shows that even with the interface recombination velocity of state-of-the-art GaAs/(Al,Ga)As$(001)$ double heterostructures, the resulting maximum lifetime would be on the order of $5$\,ns. Hence, even under the most optimistic assumptions, the internal quantum efficiency of GaAs/(Al,Ga)As NWs at this low excitation density is not expected to exceed $5 \times 10^{-4}$.

However, because of the purely bimolecular free-carrier recombination at room temperature, the radiative efficiency of GaAs NWs monotonically improves at least linearly with increasing excitation density. In fact, our measurements performed for large-signal excitation suggest an (extrapolated) internal quantum efficiency of $0.1$ at $300$\,K. Still, the evolution of the effective lifetime with temperature clearly shows that $\tau_{\text{nr}}$ is eventually approaching values close to $\tau_{\text{r}}$ at higher temperatures, and thus values lower than those obtained for the corresponding interface for $(001)$-oriented planar structures. \cite{Breuer_2011} 

In terms of applications, the GaAs/(Al,Ga)As NWs under investigation would obviously not be eligible for solar energy harvesting, taking place at the lower end of excitation densities used in the present work. However, they do seem to hold promise for applications such as light emitting or laser diodes, corresponding to the higher end of excitation densities. What is particularly attractive for light emitting diodes is the high extraction efficiency of the as-grown NW array without requiring any kind of postprocessing. 
Finally, for further improving the suitability of GaAs/(Al,Ga)As core/shell NWs for applications as light emitting devices, and to make them eligible for use as solar energy harvesters, it is crucial to identify the origin of the nonradiative recombination channel in GaAs/(Al,Ga)As core/shell NWs. In the present work, we have found that the internal quantum efficiency at low temperature does not depend on the Al content in the shell, contrary to what would be intuitively expected for interface recombination. However, further dedicated studies are required to decide whether the defect responsible is located at the interface or in the core, and to develop growth protocols to minimize the density of this defect.   


\section{Methods}
\label{sec:Exp}

\subsection{Synthesis}
The samples were fabricated using an \textsc{Oxford/VG V$80$} molecular beam epitaxy system equipped with solid sources for Ga, Al, and As. The NW growth took place on $p$-doped Si$(111)$ substrates covered by a patterned thermal oxide. Ordered arrays of GaAs NWs were selectively grown within $100 \times 100$\,µm$^2$ microfields, where holes of different sizes and distances (pitch) were previously patterned in the oxide layer by electron beam litho\-graphy. The substrate preparation prior to growth was carried out according to Ref.~\citenum{Kupers_2017}. Sample A was realized by first growing $4$-µm-long and $100$-nm-thick GaAs NWs utilizing the self-assisted vapor-liquid-solid mechanism and an As/Ga flux ratio of $3.9$ (see Ref.~\citenum{Herranz_2020} for details on the core growth). The NWs were then passivated with $30$-nm-thick Al$_{0.33}$Ga$_{0.67}$As shells grown at $500\,^{\circ}$C and an As/III flux ratio of $15$. Sample B and C were grown under the same conditions as sample A, but with a shell consisting of $30$-nm-thick Al$_{0.10}$Ga$_{0.90}$As. The NWs are unintentionally $p$-type doped with C as the predominant acceptor. 

\subsection{Microscopy}
The secondary electron micrographs shown in Fig.~\ref{fig1} were acquired in a \textsc{Hitachi S-$4800$} field-emission scanning electron microscope using an acceleration voltage of $5$\,kV and magnifications of $3500 \times$ (single NWs) and $20000 \times$ (NW arrays).

\subsection{Spectroscopy}
The NWs were investigated with three different µ-PL setups. Independently of the setup, the samples were mounted onto the coldfinger of a liquid He µ-PL cryostat allowing continuous control of the sample temperature between $10$ (nominally, see Supporting Information) and $300\,$K, and the excitation densities were controlled over six orders of magnitude by neutral density filters placed in the excitation path. The carrier temperature $T_C$ was determined by fitting the high-energy slope of PL spectra of both our planar GaAs reference samples and GaAs NWs by a single exponential. For low excitation density, the value of $T_C$ was found to saturate, yielding the actual lattice temperature of $(25 \pm 5)$\,K. The offset between the nominal and the actual temperature is due to the ambient $300$\,K radiation impinging onto the sample, which is located only a few mm beneath the cryostat window.\cite{Parma_2014_alt2}

The cw-PL data displayed in Figs.~\ref{fig1}d--h, \ref{fig2}, \ref{fig3} (the circles), and \ref{fig5}, were recorded with a \textsc{Horiba-Jobin Yvon Labram HR Evolution} Raman microscope. This setup features a confocal configuration (with a confocal hole size set to $100\,$µm for all measurements presented here) and a micro-positioning system for performing $xy$ maps as shown in Fig.~\ref{fig1}d. The samples were excited using a \textsc{Melles-Griot} $25$ \textsc{LHP} $925$ He-Ne Laser ($632.8\,$nm) focused onto the samples with an \textsc{Olympus} \textsc{MS Plan} $50\times$ objective having a numerical aperture of $0.55$. The spot size of the laser (or more accurately, the FWHM of the Gaussian beam) was directly measured by linescans of the PL intensity across single NWs as depicted in Fig.~\ref{fig1}d and amounts to $(2 \pm 0.4)$\,\textmu m. The FWHM of the detection spot (as defined by the objective's magnification and the confocal hole diameter) was the same, ensuring a uniform excitation density. The PL signal was collected by the same objective, passed though a notch filter suppressing laser stray light, and entered a spectrograph with a focal length of $80\,$cm equipped with $600\,$lines/mm grating and an LN$2$-cooled \textsc{Horiba Symphony II} CCD.

The cw-PL data represented by the squares and diamonds in Fig.~\ref{fig3} were acquired with a custom-build \textmu-PL setup in a non-confocal configuration. The samples were again excited by a \textsc{Melles Griot} $25$ \textsc{LHR} $925$ He-Ne laser ($632.8\,$nm) focused onto the samples with an \textsc{Olympus Mplan} $\times10$ microscope objective. The PL signal was collected by the same microscope objective and imaged onto the entrance slit of a $75\,$cm monochromator, passing an edge filter suppressing laser stray light. The signal was then dispersed with a $750\,$lines/mm grating and detected with an LN2-cooled \textsc{Acton} (In,Ga)As array. For determining the external quantum efficiency, a $500$\,\textmu m thick piece of \textsc{Zenith Polymer}\textregistered\ was mounted in the cryostat next to the sample. The diffusely reflected laser intensity was measured for three different laser power levels with the edge filter removed from the detection path, and a wide open entrance slit. The PL intensity was measured subsequently under exactly the same conditions for seven different laser power levels. Due to the wide open slit, the  detection area covers almost the entire NW microfield, and the PL spectra therefore contain a spurious ($e/D^0$,C$^0$) line regardless of excitation density. For determining the PL intensity, we thus considered only the $X_{\text{hh}}$ line. In addition, the measured intensities were corrected for the spectral dependence of the system response function which was determined using a calibrated halogen lamp.

The hyperspectral cathodoluminescence linescan in the inset of Fig.~\ref{fig1}g was recorded at a beam current of $0.2\,$nA by a \textsc{Gatan MonoCL$4$} system mounted to a \textsc{Zeiss Ultra$55$} analytical field-emission scanning electron microscope. An acceleration voltage of $5$\,kV was used for excitation, and the luminescence was dispersed by a $30$\,cm spectrograph equipped with a $600$\,lines/mm grating and an LN$2$-cooled \textsc{Princeton Instruments} Si CCD.

The TRPL data shown in Fig.~\ref{fig4} were acquired using a custom-build setup consisting of a \textsc{Coherent Mira} Ti:Sapphire femtosecond oscillator optically pumped by a \textsc{Coherent Verdi V$10$} Nd:YAG laser and a \textsc{Hamamatsu C$5680$} streak camera. The ultra-short laser pulses with a duration of $200\,$fs and a wavelength of $706\,$nm used for exciting the samples were focused onto the sample by a 75-mm-lens. The PL signal was collected by the same lens, dispersed by a \textsc{Jobin Yvon Spex} $1681$ $22\,$cm spectrometer and detected by the streak camera.   

\begin{acknowledgement}
The authors thank Michael Höricke, Carsten Stemmler and Claudia Herrmann for MBE maintenance, Chiara Sinito for experimental support and discussions, Anne-Kathrin Bluhm, Abbes Tahraoui, Sebastian Meister, Sander Rauwerdink, as well as Olaf Krüger, Mathias Matalla and Ina Ostermay (Ferdinand-Braun-Institut, Berlin) for technical support, Eva Monroy (CEA, France) for discussions and support with \textsc{nextnano}\texttrademark\ simulations, and Mingyun Yuan for a critical reading of the manuscript. Reference samples R$1$ and R$2$ were courteously provided by Klaus Biermann. Special thanks are due to Holger T.\ Grahn for his continuous encouragement and support. We acknowledge funding from the Alexander von Humboldt foundation (R.B.L.), the excellence cluster EXC-$2046/1$ MATH$+$ (Berlin Mathematics Research Center) of the German research foundation (O.M.), and the German Federal Ministry of Education and Research in the framework of project MILAS (J.H.).
\end{acknowledgement}


\section{Supporting information}
\subsection{Investigated samples}
Table~\ref{SI-table1} gives an overview of all samples investigated in our study. For the continuous wave-photoluminescence (cw-PL) measurements, samples A, B, C, R$1$ and R$2$ were investigated in a confocal setup and samples C and S in a home-build non-confocal setup. Time-resolved PL measurements were performed only on sample A. More details can be found in the Methods section of the main text.

\begin{table*}
\caption{Summary of the samples under investigation. $N_A^-$ denotes the ionized acceptor density in NWs (see main text for a discussion) and $p$ the hole density in layers as determined by Hall measurements at room temperature.}
\begin{tabular}{ccccc}
Sample & Type & $p/N_A^-$ (cm$^{-3}$)\\
\hline
\hline
A  & GaAs/Al$_{0.33}$Ga$_{0.67}$As NWs & ---/$<10^{16}$   \\
B  & GaAs/Al$_{0.10}$Ga$_{0.90}$As NWs &---/$<10^{16}$  \\
C & GaAs/Al$_{0.10}$Ga$_{0.90}$As NWs & ---/$<10^{16}$   \\
R$1$  & $10\,$µm-thick GaAs layer  & $4\times10^{12}$/---   \\
R$2$  & $10\,$µm-thick GaAs layer & $2\times10^{14}$/---   \\
S  & Zenith Polymer\textregistered & ---/---  \\
\hline
\hline
\end{tabular}
\label{SI-table1}
\end{table*}

\subsubsection{Comparison of optical transitions, carrier and lattice temperature}
\label{sec:comp}

Figure~\ref{SI-figure1}a depicts µ-cw-PL spectra of nanowire (NW) ensembles from sample A recorded at a nominal temperature of $10\,$K and different excitation densities. Figure~\ref{SI-figure1}b shows µ-cw-PL spectra of the reference sample R$1$ taken at comparable excitation densities. Each excitation density corresponds to a specific color of the spectrum in Figs.~\ref{SI-figure1}a,b. The lines observed for sample R$1$ are considerably narrower than the ones recorded from sample A, and we can resolve several individual transitions as shown in the inset of Fig.~\ref{SI-figure1}b. Table~\ref{SI-table2} summarizes the energy and origin of these transitions. The inset also shows the exponential high-energy tail of band-to-band recombination, from which we can deduce the carrier temperature $T_C$. For this and similar samples and for low excitation densities, $T_C$ was found to saturate at about $(30 \pm 10)$\,K, which we thus identify as the actual lattice temperature at a nominal temperature of $10$\,K. This temperature is also consistent with the comparatively high intensity of the band-to-band transition.\cite{Grilli_1992} A value of $(25\pm5)\,$K has been determined for sample A.  

\subsubsection{Mott transition in planar reference samples} 
\begin{figure}[!t]
	\includegraphics[width=\columnwidth]{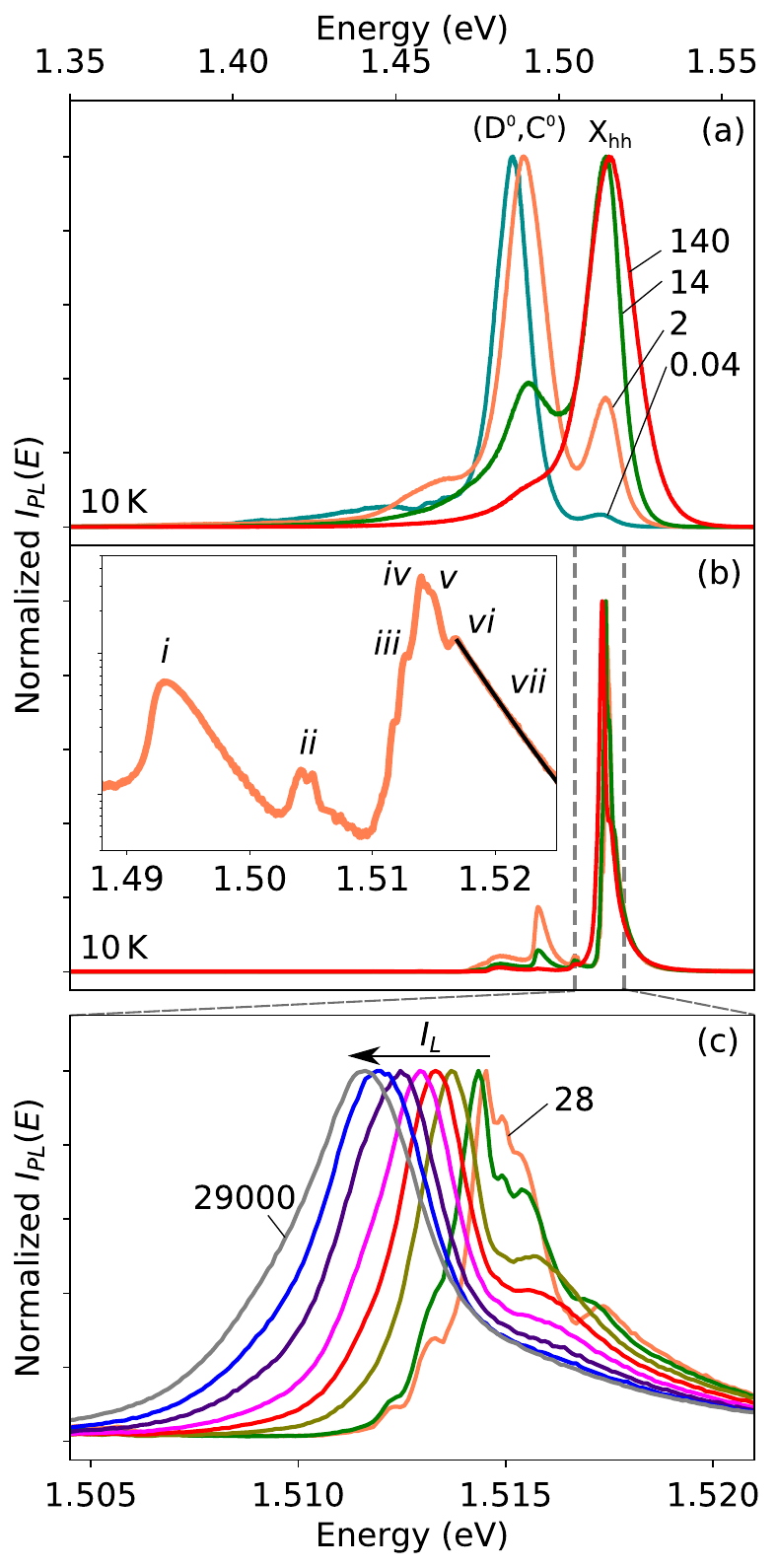}
	\caption{Normalized low-temperature ($10\,$K) µ-cw-PL spectra of (a) a NW array with a hole diameter of $90\,$nm and a pitch of $700\,$nm from sample A, and (b,c) the planar reference sample R$1$. Each spectrum is labeled with the corresponding excitation density in W\,cm$^{-2}$. The inset of (b) shows a log-lin representation of a spectrum of sample R$1$ acquired at an excitation density of $28$\,W\,cm$^{-2}$. The origin of the lines labeled as \textit{i}--\textit{vii} is summarized in Table~\ref{SI-table2}. The high energy side of the band-to-band transition is fitted with $e^{\frac{-E}{k_BT}}$ (black line).} 
	\label{SI-figure1}
\end{figure}

Figure~\ref{SI-figure1}c depicts µ-cw-PL spectra from sample R$1$ recorded in a spectral range close to the free exciton energy for a wider range of excitation densities. The spectra are observed to gradually redshift above a certain excitation density. This redshift is expected to occur after the Mott transition from excitonic to electron-hole-plasma recombination due to bandgap renormalization.\cite{Kilimann_1977,Fehrenbach_1982} A high-energy broadening indicating band-filling is not observed even for the highest excitation densities, as expected for an ideal semiconductor with a negligible density of acceptor states.\cite{Zhang_1990}. However, the excitonic transitions persist even for the highest excitation densities, which is the inevitable consequence of the nonuniform excitation of bulk-like layers.\cite{Kappei_2005}  

\begin{table}[!b]
\caption{Energy of the lines observed in the PL spectra of sample R$1$ (cf.\ Figs.~\ref{SI-figure1}b) and the underlying electronic transitions according to Ref.~\citenum{Pavesi_1994}.}
\begin{tabular}{ccccc}
Line & Energy (eV) & Transition \\
\hline
\hline
\textit{i}  & $1.4935$ & (CB,C$^0$)  \\
\textit{ii}  & $1.504$--$1.508$ & n-v lines\cite{Eaves_1984,Skolnick_1985,Beye_1985}  \\
\textit{iii}  & $1.5125$ & ($A^0$,$X_{\text{hh}}$)  \\
\textit{iv} & $1.5141$  & ($D^0$,$X_{\text{hh}}$)  \\
\textit{v}  & $1.5153$ & $X_{\text{hh}}^{n=1}$  \\
\textit{vi}  & $1.5181$ & $X_{\text{hh}}^{n=2}$    \\
\textit{vii}  & $1.5192$ & $\text{band-to-band}$    \\
\hline
\hline
\end{tabular}
\label{SI-table2}
\end{table}     
 
\begin{figure}[!t]
	\centering
	\includegraphics[width=\columnwidth]{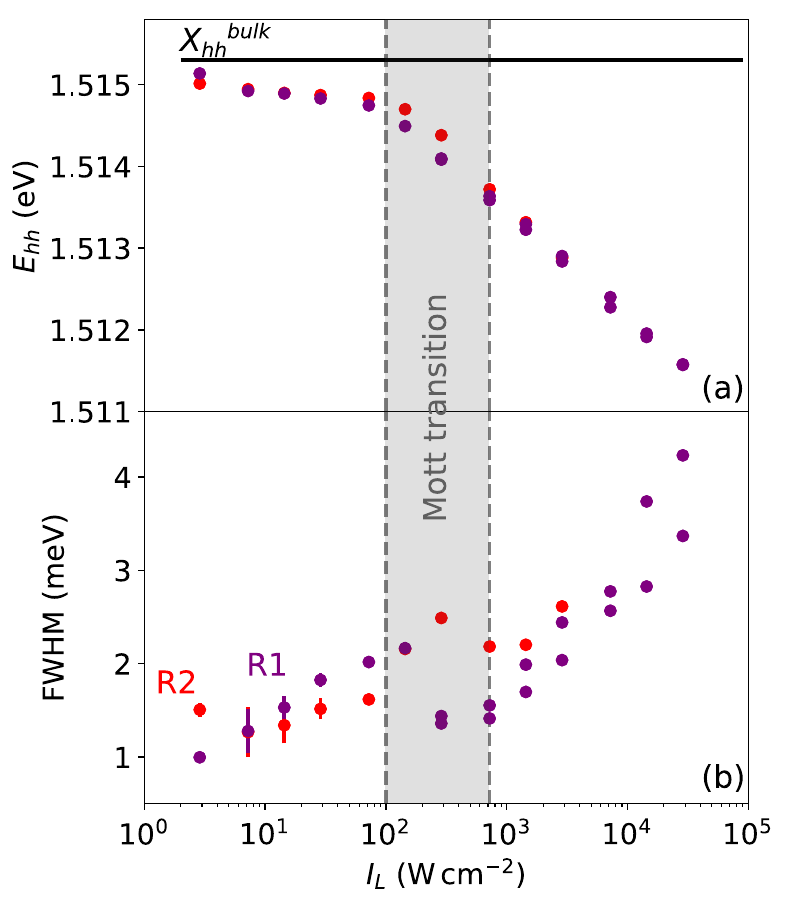}
	\caption{(a) Spectral position and (b) FWHM of the X$_{\text{hh}}$ transition of sample R$1$ (violet) and R$2$ (red) over $I_L$. The solid line at $1.5153\,$eV in (a) indicates the energy of the X$_{\text{hh}}$ line in bulk GaAs \cite{Pavesi_1994}. The region in gray indicates the gradual Mott transition from excitonic to electron-hole-plasma recombination.} 
	\label{SI-figure2}
\end{figure}

The spectral position and the width of the free exciton transition is obtained by line-shape fits integrating, for simplicity, over both lines \textit{iv} and \textit{v}. Figure~\ref{SI-figure2} shows (a) the peak position and (b) the full-width at half-maximum (FWHM) of the $X_{\text{hh}}$ transition of samples R$1$ (red symbols) and R$2$ (violet symbols) over the excitation density $I_L$. For both samples and for $I_L$ lower than $100\,$W\,cm$^{-2}$, the peak position is essentially constant, as expected for free exciton transitions,\cite{Shah_1977a,Egorov_1977,Fehrenbach_1982} and close to the X$_{\text{hh}}$ energy in bulk GaAs (indicated by the horizontal line) with the slight redshift being due to the inclusion of the donor-bound exciton in the fit. At excitation densities higher than $100\,$W\,cm$^{-2}$, the $X_{\text{hh}}$ position is seen to linearly redshift with increasing $I_L$. The FWHM of the $X_{\text{hh}}$ transition is obtained in the same way and amounts to $(1.6 \pm 0.4)$\,meV for excitation densities below $100\,$W\,cm$^{-2}$, and monotonically increases with increasing $I_L$ for excitation densities larger than $720\,$W\,cm$^{-2}$. 

As a consequence of these two observations, we place the Mott transition between $100$ and $720\,$W\,cm$^{-2}$. Contrary to the sharp Mott transition observed for the NWs (see discussions concerning Fig.~2 in the main text), the one observed here is more gradual. This gradual character of the Mott transition is primarily related to the fact that the excitation in the bulk is, in general, nonuniform.\cite{Kappei_2005}


\subsection{Reproducibility of samples and measurements}
\begin{figure}[!t]
	\centering
	\includegraphics[width=\columnwidth]{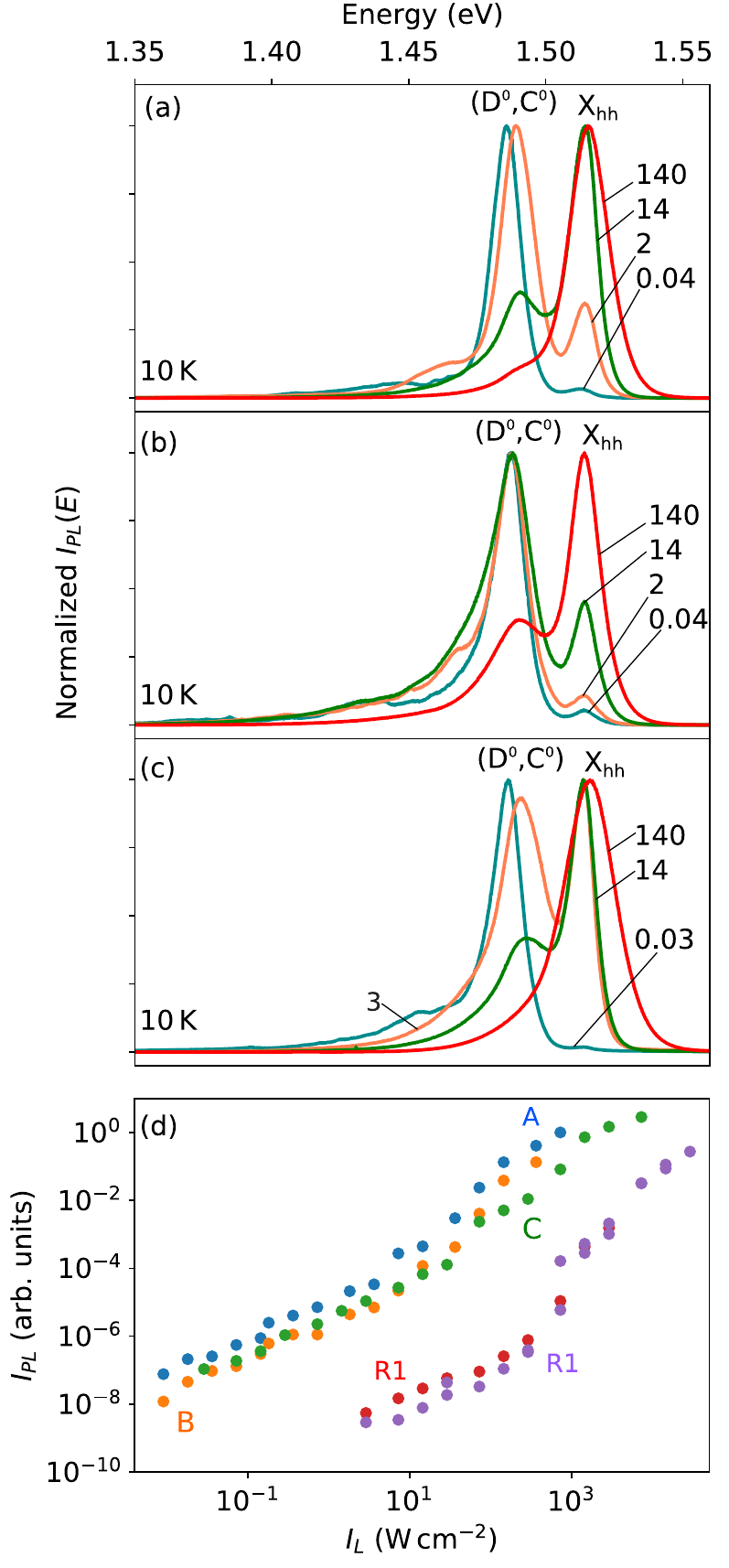}
	\caption{Normalized low-temperature ($10\,$K) µ-cw-PL spectra of NW arrays with a hole size of $90\,$nm and a pitch of $700\,$nm from sample (a) A, (b) B, and (c) C. Each spectrum is labeled with the corresponding excitation density in W\,cm$^{-2}$. (d) Integrated intensity $I_{PL}$ of the $X_{\text{hh}}$ transition of samples A (blue), B (orange), C (green), R$1$ (violet), and R$2$ (red) versus $I_L$. } 
	\label{SI-figure3}
\end{figure}

Figures~\ref{SI-figure3}a--c depict µ-cw-PL spectra of NW ensembles with a hole size of $90\,$nm and a pitch of $700\,$nm from sample A, B, and C, respectively, recorded at different excitation densities in the confocal setup. These spectra show a very low degree of polytypism, and are comparable to each other, demonstrating a high degree of reproducibility of the samples.

Figure~\ref{SI-figure3}d depicts the dependence of the spectrally integrated emission intensity $I_\text{PL}$ from samples A, B, C, R$1$ and R$2$ on excitation density $I_L$. The samples were measured side-by-side in the confocal PL setup, and samples A, B, and C are found to exhibit very similar absolute values of $I_\text{PL}$, and a very similar dependence of $I_\text{PL}$ on $I_L$. Clearly, our growth protocol is suitable for the reproducible fabrication of phase-pure GaAs/(Al,Ga)As core/shell NWs. Analogously, samples R$1$ and R$2$ are comparable as well. The PL intensity of these samples is about two orders of magnitude lower than the PL intensity emitted by the NW samples, which is due to the unpassivated surfaces of the layers, as well as to the enhanced light absorption\cite{Heiss_2014} and extraction\cite{Hauswald_2017} of NW arrays.    

\subsection{Simulations}
\label{sec:simulations}
\subsubsection{Strain components and piezoelectric potential}
\label{sec:strain}

\begin{figure*}[!t]
	\centering
	\includegraphics[width=\textwidth]{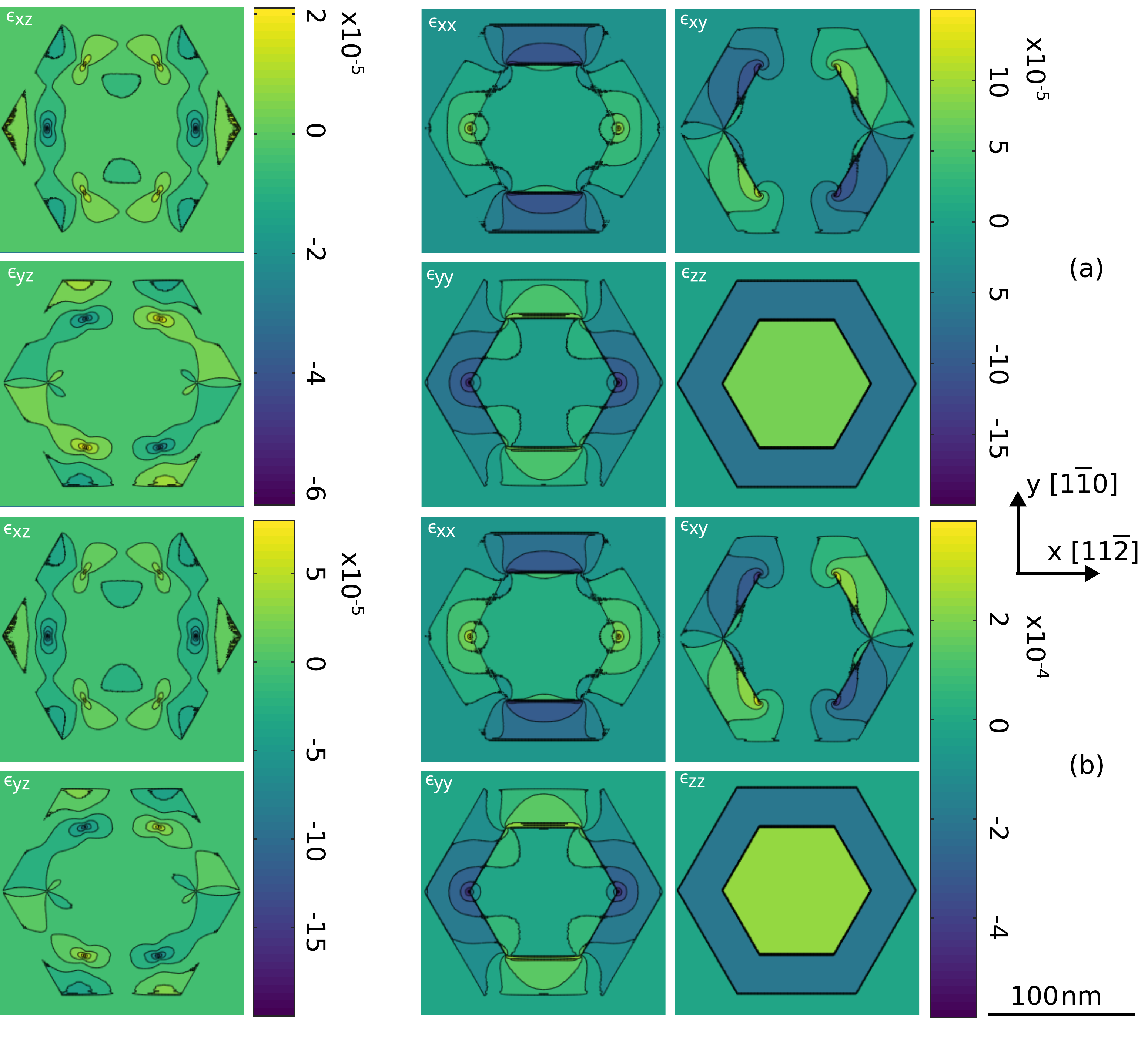}\\
	\vspace*{2mm}
	\caption{Calculated strain components of a (a) GaAs/Al$_{0.1}$Ga$_{0.9}$As and (b) GaAs/Al$_{0.3}$Ga$_{0.7}$As core/shell NW heterostructure. The diameter of the GaAs NW core along $[11\bar{2}]$ is $100\,$nm and the thickness of the (Al,Ga)As NW shell is $30\,$nm. } 
	\label{SI-figure4}
	\end{figure*}

In the following, we present an overview of the results obtained from two-dimensional ($2$D) simulations with the $8$-band $\mathbf{k \cdot p}$ formalism \cite{Marquardt_2014} as implemented within the \textsc{SPHInX} suite\cite{Boeck_2011} with parameters taken from Ref.~\citenum{Vurgaftman_2001}. Figure~\ref{SI-figure4} depicts the strain components of an (a) GaAs/Al$_{0.1}$Ga$_{0.9}$As and (b) GaAs/Al$_{0.3}$Ga$_{0.7}$As core/shell NW heterostructure. The results are consistent with previous simulations of zincblende core/shell NW heterostructures.\cite{Boxberg_2012,Glas_2015} In particular, the shear strain components $\epsilon_{xz}$ and $\epsilon_{yz}$ are found to be essentially negligible, while the shear strain component $\epsilon_{xy}$ is on the same order of magnitude as the diagonal strain components. The strain along the $z$ axis is uniform in shell and core, and the sum $(\epsilon_{xx}+\epsilon_{yy})/2$ exhibits a sixfold symmetry as reported previously\cite{Glas_2015}. Finally, the symmetry of each strain component does not depend on the Al content, but the magnitude changes linearly. 

\begin{figure}[!t]
	\includegraphics[width=\columnwidth]{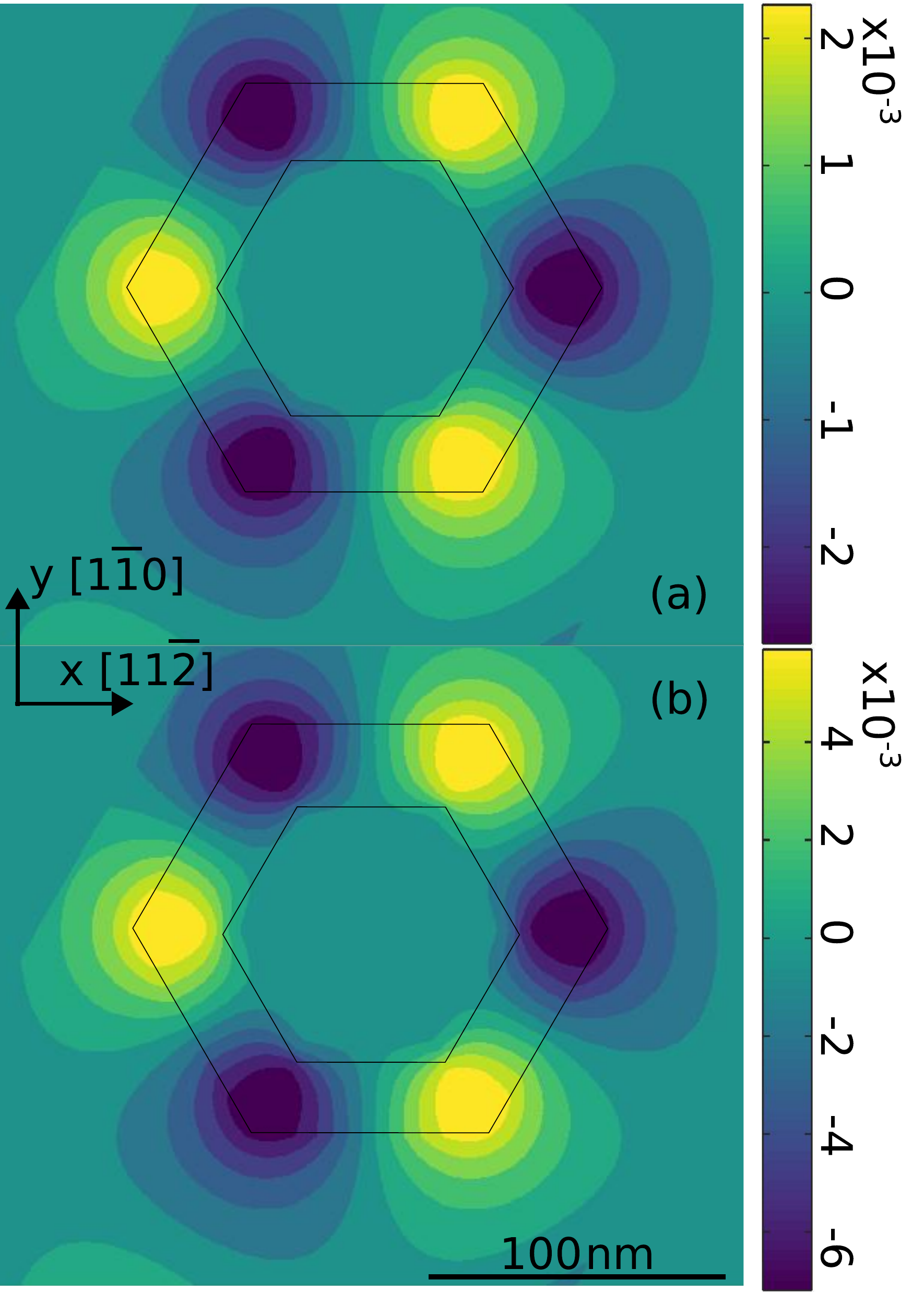}\\
	\caption{Calculated piezoelectric potential profile of an axial (a) GaAs/Al$_{0.1}$Ga$_{0.9}$As and (b) GaAs/Al$_{0.3}$Ga$_{0.7}$As core/shell NW heterostructure superimposed with a schematic drawing indicating the core/shell and shell/air interfaces. The vertical scale bar is in V. } 
	\label{SI-figure6}
\end{figure}

Figure~\ref{SI-figure6} depicts the calculated piezoelectric potential in the plane perpendicular to the NW axis of a (a) GaAs/Al$_{0.1}$Ga$_{0.9}$As and a (b) GaAs/Al$_{0.3}$Ga$_{0.7}$As core/shell NW heterostructure. For both structures, the calculated potential profiles have periodic maxima and minima with a threefold symmetry, in exact agreement with previous calculations for strained zincblende core/shell heterostructures.\cite{Boxberg_2012,Moratis_2016} While the maximum field induced by this potential remains well below $1$\,kV/cm for the Al content used in the present work, these radial piezoelectric fields must not be neglected in general. 

\begin{figure}[!t]
	\includegraphics[width=\columnwidth]{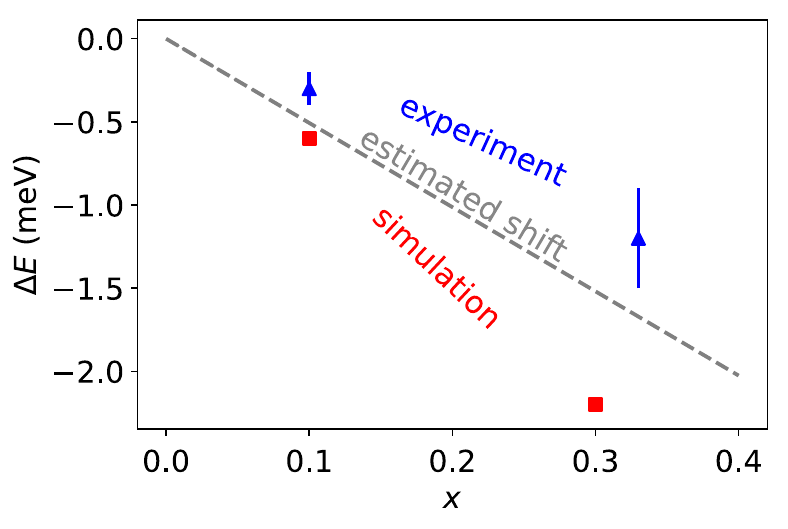}
	\caption{Experimental and calculated shifts of the transition energy for different Al content in the shell. The blue triangles depict the averaged energy shift of the measured $X_{\text{hh}}$ transition in the excitation range of $0.03$--$300$\,W\,cm$^{-2}$ with respect to the heavy-hole exciton energy in bulk GaAs ($1.5153$\,eV). The red squares show the shift of the single-particle conduction band to heavy-hole band transition taking into account both strain and the piezoelectric potential with respect to the bandgap of bulk GaAs ($1.5192$\,eV). The dashed gray line shows a simple analytical estimate taking into account only $\epsilon_{zz,C}$ as discussed in the text. 
	} 
	\label{SI-figure8}
\end{figure}

Since the GaAs core is under tensile strain, the heavy-hole valence band resides at higher energy than the light-hole one.\cite{OReilly_1989,Marquardt_2017} Hence, the experimental transition energies have to be compared to the calculated electron-to-heavy-hole transition energies. Figure~\ref{SI-figure8} depicts the measured energy shift with respect to the bulk exciton energy (blue triangles) and the calculated energy shift $\Delta E (x) = E_1 (x) - E_0 (x)$ (red squares) with respect to the bulk band gap versus the Al content. Here, $E_1$ is the single-particle transition energy calculated with \textsc{SPHInX} by taking into account the contribution from strain and the piezopotential, and $E_0$ is the bulk band gap. The experimentally obtained values for $\Delta E$ were obtained from the averaged position of the $X_{\text{hh}}$ line measured at excitation densities smaller than $300\,$W\,cm$^{-2}$ on different microfields of samples A and B. 

The calculated energy shift is notably larger than the experimental one. Since the calculated shift includes contributions due to both strain and the piezoelectric potential, we estimate the former using a simple analytical approach. Assuming that both core ($C$) and shell ($S$) share a coherent interface and thus an average (strained) lattice constant $\bar{a}$ in $z$ direction, and neglecting the difference in elastic constants between core and shell, we have:

\begin{equation}
\bar{a} = \frac{A_{C} a_{C} + A_{S} a_{S}}{ A_{C} + A_{S}}.
\end{equation}

Here, $a_{C}$ ($a_{S}$) is the unstrained lattice constant for the core (shell), and $A_{C}$ ($A_{S}$) is the cross-sectional area of the core (shell). The uniaxial strain components in the NW core $\epsilon_{zz,C}$ and in the shell $\epsilon_{zz,S}$ read:

\begin{equation}
\epsilon_{zz,C} = \frac{\bar{a}- a_{C}}{a_{C}} = \frac{A_S}{A_C+A_S}\frac{a_S-a_C}{a_C}
\label{eq:core_strain}
\end{equation}
\begin{equation}
 \epsilon_{zz,S} = \frac{\bar{a}- a_{S}}{a_{S}} = \frac{A_C}{A_C+A_S}\frac{a_C-a_S}{a_S}
 \label{eq:shell_strain}
\end{equation}
The expression deduced for the strain in the core is an approximation of the exact solution derived by \citet{Hestroffer_2010} by minimizing the strain energy in the structure. The two expressions differ by less than $0.3$\% for an Al content of $0.33$. 
The approximate solution was previously also used by \citet{Hocevar_2013} for GaAs/(Al,Ga)As core/shell NWs.

To obtain the values of $\epsilon_{zz,C}$ and $\epsilon_{zz,S}$ for the present NWs, we take the lattice constants of $a_C = a_\text{GaAs}=5.65325\,$\r{A}, $a_\text{AlAs}=5.66164\,$\r{A}, $a_S = a_{\text{Al}_{x}\text{Ga}_{1-x}\text{As}} =  x a_\text{AlAs} + (1-x) a_\text{GaAs}$,\cite{Vurgaftman_2001} where $x$ is the Al content in the shell. For simplicity, we assume a cylindrical shape of the NWs, so that the cross sectional area of the core (shell) amounts to $A_{C}=\pi r_{C}^2$ [$A_{S}=\pi (r_{NW}^2 - r_{C}^2)$]. Here, $r_{C}=50\,$nm and $r_{NW}=80\,$nm are the radii of core and the entire NW, respectively.

\begin{table}[!t]
\caption{Calculated and estimated values for $\epsilon_{\text{zz}}$ (in units of $10^{-4}$) for different Al contents [Al].}
\begin{tabular}{ccccc}
\hline
\hline
\multicolumn{1}{c}{} & \multicolumn{2}{c}{Calculated} & \multicolumn{2}{c}{Estimated} \\
\hline
\multicolumn{1}{c}{[Al]} & core & shell & core & shell \\
\hline
$0.1$  & $0.84$ & $-0.54$ & $0.90$ & $-0.58$ \\
$0.3$  & $2.53$ & $-1.61$ & $2.72$ & $-1.74$ \\
\hline
\hline
\end{tabular}
\label{SI-table3}
\end{table}

Table \ref{SI-table3} compares the values for $\epsilon_{zz,C}$ and $\epsilon_{zz,S}$ obtained by \textsc{SPHInX} and the simple estimate outlined above. Despite the simplifying assumptions, the estimated values deviate by less than $10$\% from the calculated ones. For a purely uniaxial tensile strain along the $[111]$ direction, the band gap shrinks linearly with strain at a rate of $56$\,meV/\%.\cite{Senichev_2018} The dashed line in Fig.~\ref{SI-figure8} shows the corresponding dependence versus the Al content. Despite the slightly higher strain values compared to the simulation (cf.\ Table~\ref{SI-table3}), the trend predicted by our simple estimate is actually closer to the experimental data than the $2$D simulation. The reason for the larger redshift in the simulation is the fact that the single-particle states are affected by the radial piezoelectric potential depicted in Fig.~\ref{SI-figure6}, while the exciton dominating the experimental spectra is not. Hence, the redshift of the $X_{\text{hh}}$ line can be fully attributed to the tensile strain exerted on the GaAs core by the respective (Al,Ga)As shell. 

\subsubsection{Radial electric fields due to surface band bending}
Next, we address the magnitude of radial electric fields in the GaAs core arising from surface band bending induced by charge transfer from bulk acceptor to surface states, and their consequences on carrier recombination. 

\begin{figure}[!t]
	\centering
	\includegraphics[width=\columnwidth]{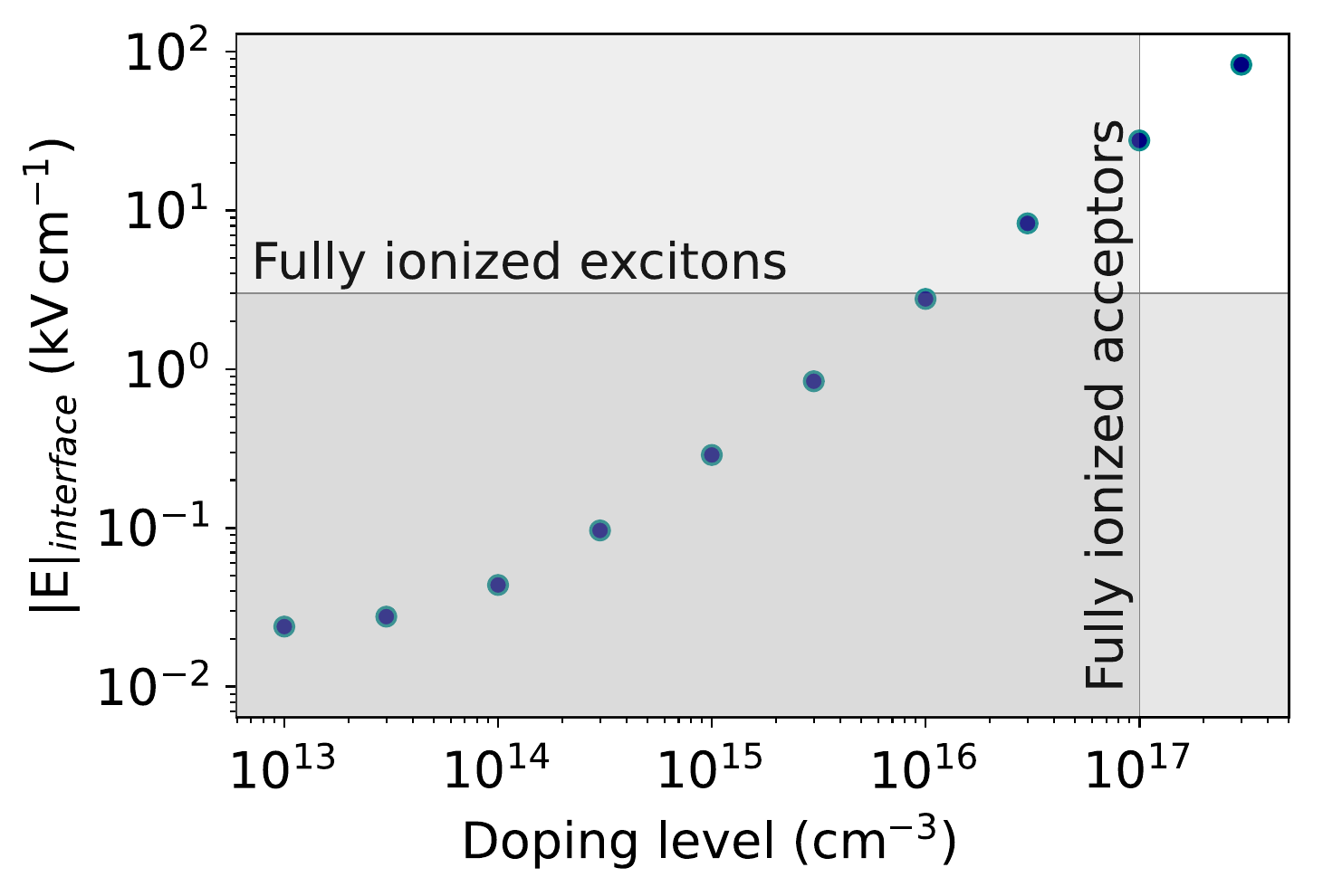}
	\caption{Magnitude of the electric field in the GaAs core close to the GaAs/Al$_{0.3}$Ga$_{0.7}$As interface versus the concentration of shallow acceptors calculated with the Poisson solver implemented in \textsc{nextnano}\texttrademark. The horizontal line at $3\,$kV\,cm$^{-1}$ indicates the critical field above of which, in GaAs, all excitons are ionized. The acceptors are fully ionized up to a doping level of $10^{17}\,$cm$^{-3}$ as indicated by the vertical line.	
	} 
	\label{SI-figure9}
\end{figure}

Figure~\ref{SI-figure9} depicts the magnitude of the electric field in the GaAs core close to the core/shell interface as a function of acceptor concentration, calculated with the $2$D Poisson solver in \textsc{nextnano}\texttrademark\ of a GaAs/Al$_{0.3}$Ga$_{0.7}$As core/shell NW at $10\,$K. The strength of the internal radial electric field stays below the exciton ionization threshold for doping levels up to $10^{16}\,$cm$^{-3}$. 
Additionally, these simulations show that the GaAs NW core is fully depleted for doping levels up to $10^{17}\,$cm$^{-3}$.
The magnitude of the electric fields associated with surface band bending depends not only on acceptor concentration, but also on the diameter of the NWs. \citet{Geijselaers_2018} reported that the variation of the internal electric fields with diameter can be extracted from cw-PL spectra as a systematic shift of the emission energy. The band bending would also lead to a superposition of transitions with different energies and thus potentially influence the FWHM of the PL emission. 

\begin{figure}[!t]
	\centering
	\includegraphics[width=\columnwidth]{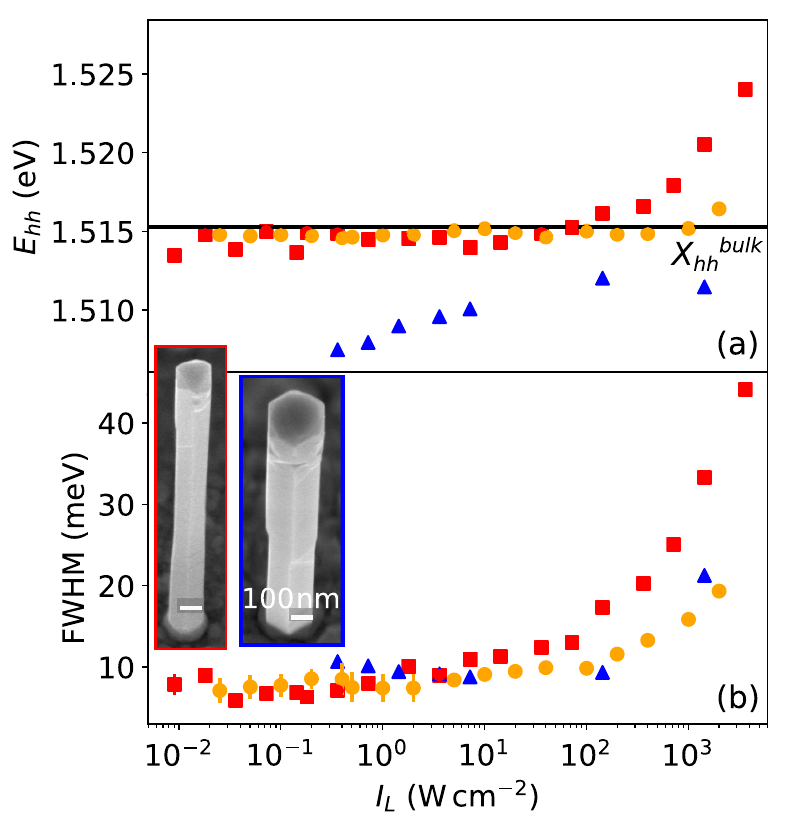}
	\caption{(a) $E_{hh}$ and (b) FWHM of the highest energy line in the low-temperature ($10\,$K) PL spectra of NW arrays with a hole size of $90\,$nm and a pitch of $700\,$nm from sample A (red squares) and B (orange circles), and a hole size of $40\,$nm and a pitch of $1000\,$nm from sample D (blue triangles) versus $I_L$. The inset in 
	(b) displays secondary electron micrographs of representative NWs from samples A and D. The value of both scale bars in the secondary electron micrographs is $100\,$nm. } 
	\label{SI-figure10}
\end{figure}

Figure~\ref{SI-figure10} depicts both the transition energy (a) and the width (b) of the highest energy line in the PL spectra of NW arrays from three samples, namely, the arrays with a hole size of $90\,$nm and a pitch of $700\,$nm from sample A and B, and an array with a hole size of $40\,$nm and a pitch of $1000\,$nm from sample D. The latter sample was grown under the same conditions as sample A, with the only difference that the core diameter has been increased by depositing an additional GaAs shell with a thickness of $25\,$nm (resulting in an increased core diameter by $50\,$nm) was grown prior to the growth of the $30\,$nm thick Al$_{0.3}$Ga$_{0.7}$As shell. 
The increased diameter of the NWs in sample A and D can be seen in secondary electron micrographs of representative single NWs shown in the inset of Fig.~\ref{SI-figure10}b.

While samples A and B exhibit a constant transition energy only slightly redshifted by strain, sample D shows a very different behavior: the redshift is much stronger, particularly at low $I_L$, and decreases gradually with increasing $I_L$. Since the strain in the core is reduced for sample D compared to samples A and B, this finding suggests that the radial electric field in sample D is of sufficient strength to dissociate excitons, resulting in recombination of spatially separated electron-hole pairs with an energy redshifted by a radial Stark effect.    


 

\subsection{Recombination model for fitting $\boldsymbol{I_{PL}(I_L)}$ and $\boldsymbol{\eta_\text{int}(I_L)}$}
We start from a coupled system of rate equations for the densities of excitons, electrons and holes, shallow acceptors, and nonradiative Shockley-Read-Hall centers. Following Refs.~\citenum{Brandt_1995a,Brandt_1998a}, we can summarize this equation system into one effective equation reading
\begin{equation}
G-B \Delta n \left( \hat{p}_0+ \Delta p \right)-\frac{\Delta n \Delta p}{\tau_p \Delta n + \tau_n \Delta p} = 0,
\label{eq_ModelI}
\end{equation}
with the first term accounting for steady-state generation with rate $G$, the second for radiative, and the third for nonradiative recombination. 
This expression is mathematically equivalent to Eq.~5 in Ref.~\citenum{Brandt_1996}, but with $\hat{p}_0$ denoting not the background carrier concentration (which is essentially zero in our fully depleted NWs), but representing the monomolecular recombination due to shallow acceptors. In all what follows, symbols not explicitly defined have their usual meaning. 

In the second term of Eq.~\ref{eq_ModelI}, $B=b_r + \sigma_x \gamma_x$ is the effective radiative recombination coefficient including the contributions of free carriers and excitons. The former \cite{Brandt_1998b} is given by 
\begin{equation}
b_r=\frac{\Gamma}{N_x}
\end{equation}
with the dipole transition rate
\begin{equation}
\Gamma = \frac{n E_G e^2 \mid p_{cv} \mid^2}{3 \pi m_0^2 \hbar^2 c^3 \epsilon_0}
\end{equation}
and the total effective density of states
\begin{equation}
 N_{T}=\frac{1}{\sqrt{2}} \left( \frac{(m_e+m_{hh}) k_B T}{\pi{} \hbar^2} \right) ^{3/2}.
\end{equation}
In the contribution of excitons, 
\begin{equation}
\sigma_x = \frac{2}{N_{cv}} e^{E_x/k_B T}
\end{equation}
is the scattering volume for free carriers into the exciton state with the reduced density of states
\begin{equation}
 N_{cv}=\frac{1}{\sqrt{2}} \left( \frac{\mu k_B T}{\pi{} \hbar^2} \right) ^{3/2}
\end{equation}
and $\gamma_x$ is the exciton's radiative rate. 

The monomolecular radiative term is proportional to 
\begin{equation}
\hat{p}_0=\frac{b_A}{B} (N_A - \Delta n + \Delta p ) \frac{\Delta p}{\Delta n}
\end{equation}
with the recombination coefficient for band-to-acceptor recombination \cite{Bebb_1972aa}
\begin{equation}
	b_A = \frac{2^5 \sqrt{2} \hbar n E_G e^2 \mid p_{CV} \mid ^2 }{4 \pi m_0^2 c^3 \epsilon_0 (m_h E_A)^{3/2}} \mathcal{I_A},
\end{equation}
where
\begin{equation}
	\mathcal{I_A} = 2 \pi \left( \frac{\beta}{\pi} \right)^{3/2} \int_{0}^{\infty} \frac{\sqrt{x} e^{-\beta x}}{(1 + x)^4}  dx
\end{equation}
with
\begin{equation}
	\beta = \frac{m_h E_A}{m_e k_B T}.
\end{equation}

In the third term, the electron and hole capture times are given by
\begin{equation}
\tau_n = \frac{1}{b_n N_C}~\text{and}~\tau_p = \frac{1}{b_p N_C},	
\end{equation}
respectively, where $b_n$ and $b_p$ are the respective capture coefficients and $N_C$ is the density of Shockley-Read-Hall centers.

In addition to Eq.~\ref{eq_ModelI}, we have to explicitly consider charge neutrality:
\begin{equation}
\Delta n + N_A^- = \Delta p + N_C^+
\label{eq_chargeneutrality}
\end{equation}
where
\begin{equation}
	N_A = N_A^- + N_A^0~\text{and}~N_C= N_C^+ + N_C^0.
\end{equation}
Note that at $10$\,K, the quasi-equilibrium between neutral (occupied) and ionized (empty) acceptor states implies that  
\begin{equation}
		N_A^- = \frac{N_A}{1 + \sigma_A \Delta p} \approx 0,	
\end{equation}
with 
\begin{equation}
	\sigma_A = \frac{2}{N_V} e^{E_A/k_BT},
\end{equation}
where 
\begin{equation}
N_V = \frac{1}{\sqrt{2}} \left( \frac{m_h k_B T}{\pi \hbar^2} \right)^{3/2}
\end{equation}
is the valence band effective density of states.

Equations \ref{eq_ModelI} and \ref{eq_chargeneutrality} define model II as investigated in detail in Ref.~\citenum{Brandt_1996}. When assuming $\Delta n = \Delta p$, we obtain model I as
\begin{equation}
G - B \Delta n \left( \frac{b_A}{B} N_A  + \Delta n \right) - \frac{\Delta n}{\tau_n + \tau_p} = 0,
\label{eq_ModelII}
\end{equation}
which is mathematically equivalent to the situation considered in Refs.~\citenum{Brandt_1995a,Brandt_1995b}.

All temperature-dependent quantities have been calculated assuming a lattice temperature of $25$\,K (see Sec.~\nameref{sec:comp}).





\bibliography{references}

\end{document}